\documentclass[twocolumn]{aastex61}
\hypersetup{
  pdfinfo={Title={Matsuno2017b.pdf},Author={Tadafumi Matsuno}}}
\begin{document}
\title{High-resolution spectroscopy of extremely metal-poor stars from SDSS/SEGUE. III. Unevolved Stars with $[\mathrm{Fe/H}]\lesssim -3.5$
\footnote{Based on data collected with the Subaru Telescope, which is operated by the National Astronomical Observatory of Japan.}
}
\author{Tadafumi Matsuno}
\affiliation{Department of Astronomical Science, School of Physical Sciences, SOKENDAI (The Graduate University for Advanced Studies), 2-21-1 Osawa, Mitaka, Tokyo 181-8588, Japan}
\email{tadafumi.matsuno@nao.ac.jp}
\affiliation{National Astronomical Observatory of Japan (NAOJ), 2-21-1 Osawa, Mitaka, Tokyo 181-8588, Japan}

\author{Wako Aoki}
\affiliation{National Astronomical Observatory of Japan (NAOJ), 2-21-1 Osawa, Mitaka, Tokyo 181-8588, Japan}
\affiliation{Department of Astronomical Science, School of Physical Sciences, SOKENDAI (The Graduate University for Advanced Studies), 2-21-1 Osawa, Mitaka, Tokyo 181-8588, Japan}

\author{Timothy C. Beers}
\affiliation{Department of Physics and JINA Center for the Evolution of the Elements, University of Notre Dame, Notre Dame, IN 46556, USA}

\author{Young Sun Lee}
\affiliation{Department of Astronomy and Space Science, Chungnam National University, Daejeon 34134, Korea}

\author{Satoshi Honda}
\affiliation{Nishi-Harima Astronomical Observatory, Center for Astronomy, University of Hyogo, 407-2 Nishigaichi, Sayo-cho, Sayo, Hyogo 679-5313, Japan}

\begin{abstract}

We present elemental abundances for eight unevolved extremely metal-poor
stars with $T_{\rm eff}>5500\,\mathrm{K}$, among which seven have
$[\mathrm{Fe/H}]<-3.5$. The sample is selected from the Sloan Digital Sky
Survey / Sloan Extension for Galactic Understanding and Exploration
(SDSS/SEGUE), and our previous high-resolution spectroscopic follow-up
with the Subaru Telescope (Aoki et~al.). Several methods to derive
stellar parameters are compared, and no significant offset in the
derived parameters is found in most cases. From an abundance analysis
relative to the standard extremely metal-poor star G~64--12, an average
Li abundance for stars with $[\mathrm{Fe/H}]<-3.5$ is $A(\mathrm{Li})
=1.90$, with a standard deviation of $\sigma =0.10$~dex.
This result confirms that lower Li abundances are found at lower
metallicity, as suggested by previous studies, and demonstrates that the
star-to-star scatter is small. The small observed scatter could be a
strong constraint on Li-depletion mechanisms proposed for explaining the
low Li abundance at lower metallicity. Our analysis for other elements
obtained the following results: i) A statistically significant scatter
in $[\mathrm{X/Fe}]$ for Na, Mg, Cr, Ti, Sr, and Ba, and an apparent
bimodality in $[\mathrm{Na/Fe}]$ with a separation of $\sim 0.8\,
\mathrm{dex}$, ii) an absence of a sharp drop in the
metallicity distribution, and iii) the existence of a CEMP-$s$ star at
$[\mathrm{Fe/H}]\simeq -3.6$ and possibly at $[\mathrm{Fe/H}]\simeq-4.0$, which may provide a
constraint on the mixing efficiency of unevolved stars during their
main-sequence phase.
\end{abstract}

\section{Introduction\label{sdsssecint}}

Extremely metal-poor (EMP; [Fe/H] $< -3.0$) stars provide chemical
information on the Universe at a unique phase of its evolution.
Precise cosmic microwave background (CMB) measurements from space
constrain the conditions at the time of the Big Bang \citep[e.g.,
][]{PlanckCollaboration2016}, whereas observations of galaxies across
a wide range of redshift trace galaxy evolution over cosmic time
\citep[e.g.,][]{Madau2014}. However, in order to connect galaxy formation with
the Big Bang, understanding of the formation and evolution of 
first-generation stars is indispensable. Since the chemical abundances of
EMP stars are not generally affected by nucleosynthesis processes
other than the Big Bang and the supernovae explosions of the first stars, they
can fill the gap between observations of the CMB and those of
later-forming galaxies.

Stellar Li abundances deliver uniquely important information, since Li
is the only element (beyond H and He) that is synthesized in the Big Bang 
to a significant degree and can be
measured in the atmospheres of many EMP stars. Although the constant Li
abundance found in metal-poor turn-off stars was formerly regarded as a
constraint on Big Bang nucleosynthesis \citep{Spite1982,Spite1982a},
the ``Li plateau'' value turned out to stand in contradiction to the Li
abundance predicted by Big Bang nucleosynthesis models based on the
recent CMB observations \citep{Coc2004,Cyburt2016}. Theoretical trials
invoking Li-depletion mechanisms in the formation and evolution of
low-mass metal-poor stars have attempted to explain this discrepancy
\citep[e.g.,][]{Richard2005,Piau2006,Fu2015}. One difficulty is
reproducing the small observed scatter in Li abundances for metal-poor
turn-off stars with $-2.5\lesssim[\mathrm{Fe/H}]\lesssim -1.6$. In
addition, recent observations demonstrate that the plateau breaks down
below $[\mathrm{Fe/H}]\sim -2.5$, and no star has Li abundance
comparable to the plateau below $[\mathrm{Fe/H}]=-4.0$ \citep[e.g.,
][]{Ryan1996,Ryan1999,Bonifacio2007,Frebel2008,Aoki2009,Sbordone2010,
Caffau2011,Hansen2014,Bonifacio2015,Li2015a}. 

The key stellar metallicity occurs below $[\mathrm{Fe/H}]\sim -3.0$,
especially $\lesssim -3.5$. Stars with $[\mathrm{Fe/H}]\sim -3.5$ bridge
the Spite Plateau stars and ultra metal-poor (UMP;
$[\mathrm{Fe/H}]<-4.0$) stars, all of which exhibit low lithium
abundances. However, the current sample size of turn-off stars with
$-4.0<[\mathrm{Fe/H}]<-3.5$ with available Li measurements is still
small, $\sim 10$ according to the SAGA database
\citep{Suda2008,Suda2011,Yamada2013}.

One of the reasons for the small sample size is the rarity of EMP stars.
Another is the difficulty of deriving precise abundances for warm EMP
stars. In addition to their extremely low metallicity, the relatively
high temperatures of main-sequence turn-off stars weaken their
absorption lines; higher signal-to-noise ratios for such stars are
required for precise abundance measurements.

The purpose of this study is to determine chemical abundances, including
Li, for turn-off stars with $[\mathrm{Fe/H}]<-3.5$. We have obtained
high-resolution, high--signal-to-noise ratio spectra with the Subaru
Telescope for eight warm EMP stars ($T_{\rm eff}>5500\,
\mathrm{K}$, $[\mathrm{Fe/H}]<-3.0$) previously considered by
\citet{Aoki2013}; seven of the eight have $[\mathrm{Fe/H}]<-3.5$. The
relatively narrow range of stellar parameters among the sample enable a
high-precision differential abundance analysis. Hence, besides
understanding the nature of possible Li-depletion mechanisms, these
targets are useful for investigation of other elemental abundances for
EMP stars. In addition, since our targets have not yet reached the
red-giant stage of evolution, we can examine possible abundance changes
caused by first dredge-up (e.g., \citealt{Spite2006}) from a comparison
of the chemical abundances of our targets with those of red giants
reported in the literature.

This paper is organized as follows. Details of target selection and
observation are described in Section \ref{sdsssecobs}. In Section
\ref{sdsssecparam}, we compare several methods to derive stellar
parameters for the eight targets. In addition, we also determine stellar
parameters for two bright EMP stars (G~64--12 and LP~ 815--43) with
parallaxes measured by the Gaia satellite. The abundance analysis and
its results are described in Section \ref{sdsssecchem}. After presenting
an interpretation of the results in Section \ref{sdsssecdis}, we
summarize our conclusions in Section \ref{sdsssecsum}.

\section{Observations and reduction\label{sdsssecobs}}

The targets in our present study are selected from \citet{Aoki2013}, who
reported the results of abundance analysis of snap-shot high-resolution
spectroscopy for 137 metal-poor candidates discovered by the Sloan
Digital Sky Survey \citep[SDSS; ][]{York2000} and the Sloan Extension
for Galactic Understanding and Exploration \citep[SEGUE; ][]{Yanny2009}.
We have obtained new higher-quality spectra for eight targets with the
High Dispersion Spectrograph on the Subaru Telescope
\citep{Noguchi2002}. The spectral resolution is $R=60,000$ with $2\times
2$ CCD binning; the wavelength coverage is $4000-6800\,\mathrm{\AA}$.
Details of the observations are provided in Table \ref{tabobjlst}.
Hereafter, object names are shown using abbreviations, e.g., SDSS~J0120--1001 for SDSS~J012032.63--100106.5.
Although the spectrum of one of our targets, SDSS~J1424+5615, has been analyzed in
\citet{Matsuno2017}, we re-analyze it in this study.

The data are reduced in a standard manner using the IRAF\footnote{IRAF
is distributed by the National Optical Astronomy Observatory, which is
operated by the Association of Universities for Research in Astronomy,
Inc. under cooperative agreement with the National Science Foundation.}
echelle package, including bias correction, flat fielding, scattered
light subtraction, extraction of spectra, and wavelength calibration
using Th arc lines. The signal-to-noise ratios per $1.1\,
\mathrm{km\,s^{-1}}$ pixel 
around $6708\,\mathrm{\AA}$ and per $1.5\,\mathrm{km\,s^{-1}}$
 around $4877\,\mathrm{\AA}$ (after re-binning) are estimated from the
standard deviation of the continuum level 
around the Li I doublet at $6708\,
\mathrm{\AA}$. Heliocentric radial velocities ($v_r$) are estimated from
Fe lines. Typical uncertainties in $v_r$ are $\sim 1\,\mathrm{km\,
s^{-1}}$. All but one target shows no significant radial velocity changes
from \citet{Aoki2013}. The exception is SDSS~J2349+3832, for which our
radial velocity is larger than at the epoch of August 22, 2008 by $3.1\,
\mathrm{km\,s^{-1}}$. 

We also analyze the spectra of two bright EMP main-sequence turn-off stars,
G~64--12 ([Fe/H] $= -3.38$) and LP~815--43 ([Fe/H] $= -2.96$). The spectrum of
G~64--12 was taken on December 22, 2002 with $R\sim 90,000$ and $S/N
\sim 650$ at $6708\,\mathrm{\AA}$ ($S/N \sim454$ at $4880\,\mathrm{\AA}$)
 \citep{Aoki2009}. The spectrum of LP~815--43, which was taken
from the Subaru archive SMOKA \citep{Baba2002}, was originally obtained
on May 18, 2005 with $R\sim 90,000$ and $S/N \sim 260$
 at $6708\,\mathrm{\AA}$ ($S/N \sim142$ at $4880\,\mathrm{\AA}$). 
Both stars are
included in the first data release of the Gaia satellite
\citep{GaiaCollaboration2016, GaiaCollaboration2016a}, which allows us
to obtain an independent determination of their surface gravities.

\begin{deluxetable*}{lrrrrrr}
  \tablecaption{Observation Log for the SDSS Sample \label{tabobjlst}}
  \tablehead{
\colhead{Object Name\tablenotemark{a}}& 
\colhead{Observing run}& \dcolhead{V} &\colhead{Exp. time}  &\colhead{$S/N$} &\colhead{$S/N$} &
\colhead{ $v_r$} \\ 
 \colhead{} &\colhead{(yyyy-mm-dd)} & \colhead{(mag)} &
\colhead{(s)} & \colhead{ $@4877\,\mathrm{\AA}$}& \colhead{ $@6708\,\mathrm{\AA}$} & \colhead{($\mathrm{km\,s^{-1}}$)} }
\startdata
SDSS J012032.63--100106.5 & 2009-09-10,12       &  16.55& 14490  & 34     &  66    & -58.6 \\
SDSS J103649.93+121219.8 & 2009-11-24           &  15.50& 6514   & 37     &  50    & -33.8 \\
SDSS J142441.88+561535.0 & 2009-06-28,29,07-01  &  15.70& 13800  & 76     &  85    & 0.0  \\
SDSS J152202.09+305526.3 & 2009-06-28,29        &  16.42& 21600  & 60     &  69    & -354.8 \\
SDSS J164005.30+370907.8 & 2009-06-28,29        &  15.55& 14400  & 95     &  95    & -51.6 \\
SDSS J200513.48--104503.2 & 2009-09-12          &  16.78& 12000  & 31     &  40    & -55.8  \\ 
SDSS J230959.55+230803.0 & 2009-11-24           &  17.02& 12000  & 20     &  36    & -307.0 \\
SDSS J234939.71+383217.8 & 2009-09-10,12        &  16.80& 19200  & 38     &  60    & -84.2 \\
  \enddata  
\tablecomments{Objects IDs given by SDSS can be found in \citet{Aoki2013}.}
\tablenotetext{a}{The object name is an abbreviation, and is used throughout the rest of this paper, e.g., SDSS~J0120--1001 for SDSS~J012032.63--100106.5.}
\end{deluxetable*}

\section{Stellar atmospheric parameters\label{sdsssecparam}}
\subsection{Methods}

In order to establish the most reliable method to derive stellar
parameters for EMP turn-off stars, we apply four
methods, i) analysis of Balmer-line profiles, ii) spectroscopic analysis
of Fe lines, iii) the SEGUE Stellar Parameter Pipeline \citep[SSPP;
][]{Lee2008,Lee2008a,AllendePrieto2008}, and iv) colors (only for
$T_{\rm eff}$), and compare the results. Each method is briefly
described below.

\subsubsection{Balmer-Line Profiles}

Balmer lines of hydrogen are prominent in spectra of warm stars. Their
profiles, especially the width of the wings, are sensitive to effective
temperature. Contamination arising from metallic absorption lines in the
profiles of Balmer lines is insignificant in EMP stars, with the
exception of the H$\gamma$ line in carbon-enhanced metal-poor (CEMP)
stars, which can be impacted by the presence of the CH $G$-band
molecular feature. Our procedure is essentially the same as the method
of \citet{Barklem2002}, and described in \citet{Matsuno2017}. We here
briefly summarize our approach, focusing on differences from the
previous work. 

Careful continuum placement is essential in the analysis of Balmer lines
with broad profiles. We estimate the continuum level by interpolating
across the blaze functions of adjacent orders containing these lines.

Models of Balmer-line profiles are taken from interpolation of the grid by
\citet{Barklem2002} \footnote{http://www.astro.uu.se/\%7ebarklem/}.
Atmospheres with $[\mathrm{Fe/H}]=-3$ and $[\mathrm{\alpha/Fe}] = +0.4$ are assumed
throughout the analysis. While the H$\beta$ line is only sensitive to
$T_{\rm eff}$, H$\alpha$ is also dependent on surface gravity, $\log g$. Hence, we
determine $T_{\rm eff}$ from the H$\beta$ line first, assuming $\log
g=4.0$, and then determine $\log g$ from the H$\alpha$ line. We iterate
the estimates until the set of $(T_{\rm eff},\, \log g)$ reaches
convergence (usually less than three times). Once we obtain the best-fit
spectrum from the H$\beta$ and H$\alpha$ fitting procedure, we remove
possible effects of cosmic rays and absorption lines from the observed
spectrum by masking pixels that deviate from the best-fit spectrum by more than
$2.5 \sigma$. We also modify the fitting region to only include the line
wings, defined as the regions where the normalized flux of
the best-fit model is between 0.7 and 0.9. We then repeat the fitting until
convergence is achieved (usually less than five times). 

Errors in our procedure are dominated by uncertainty of the continuum
placement. An error of $0.5\,\%$ in the continuum placement for our
sample stars is estimated by applying the interpolating procedure to the
orders that contain no broad absorption features. We estimate its effect
on $T_{\rm eff}$ and $\log g$ by analyzing the spectra whose continuum
level is artificially shifted by $0.5\,\%$. In addition, since the
estimate of surface gravity is dependent on the assumed $T_{\rm eff}$, we
calculated uncertainties of $\log g$ as follows:
\begin{equation}
\sigma_{\log g}^2 = (\delta \log g)^2 + (\frac{\partial \log g}{\partial T_{\rm eff}}\delta T_{\rm eff})^2,
\end{equation}
where $\delta X$ represents the uncertainties caused by
continuum-placement errors, $\sigma_X$ is the total uncertainties, and X
denotes either $\log g$ or $T_{\rm eff}$.
Since $\log g$ does not affect $T_{\rm eff}$ estimates, we adopt $\sigma_{T_{\rm eff}} = \delta T_{\rm eff}$.
The covariance can be found in a similar manner:
\begin{equation}
\sigma_{T_{\rm eff} \log g} = \frac{\partial \log g}{\partial T_{\rm eff}}\sigma_{T_{\rm eff}}^2.
\end{equation}
Since covariances contribute to the total errors of our derived
abundances, they need to be taken into account.

\begin{figure*}
  \plottwo{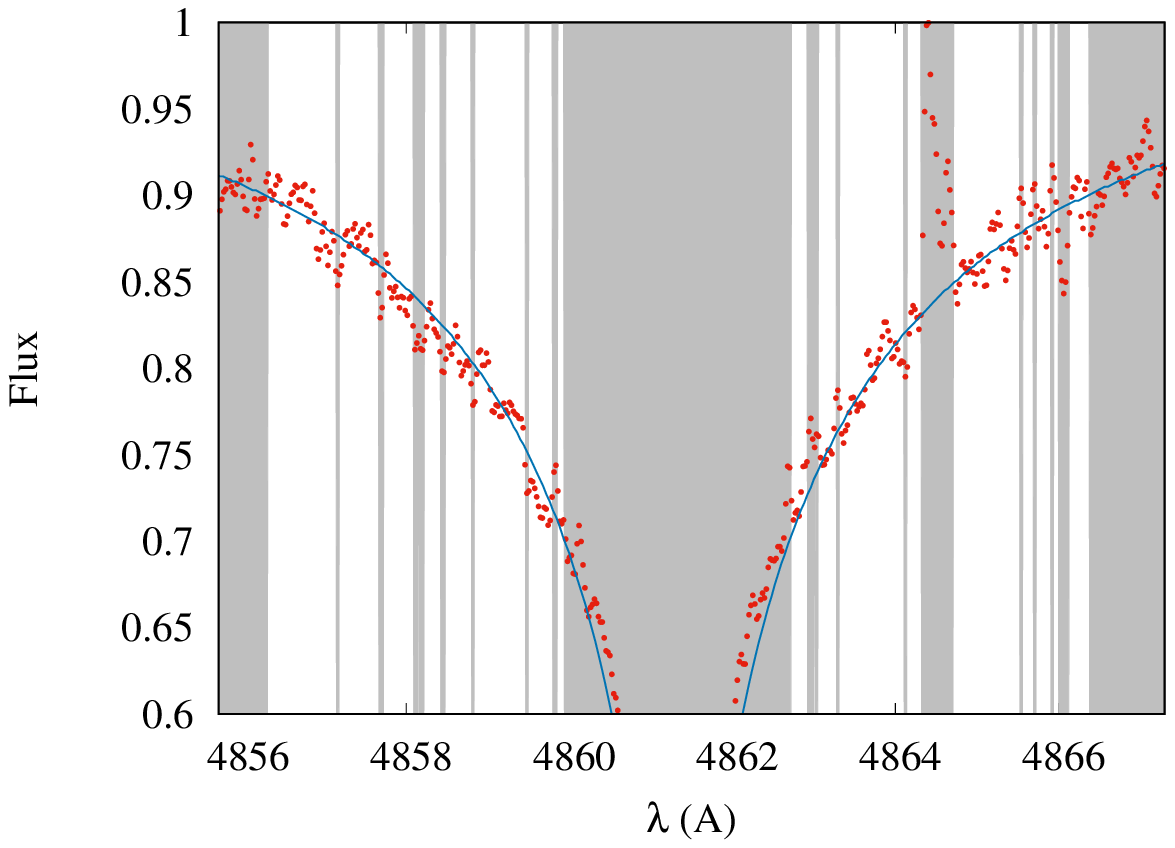}{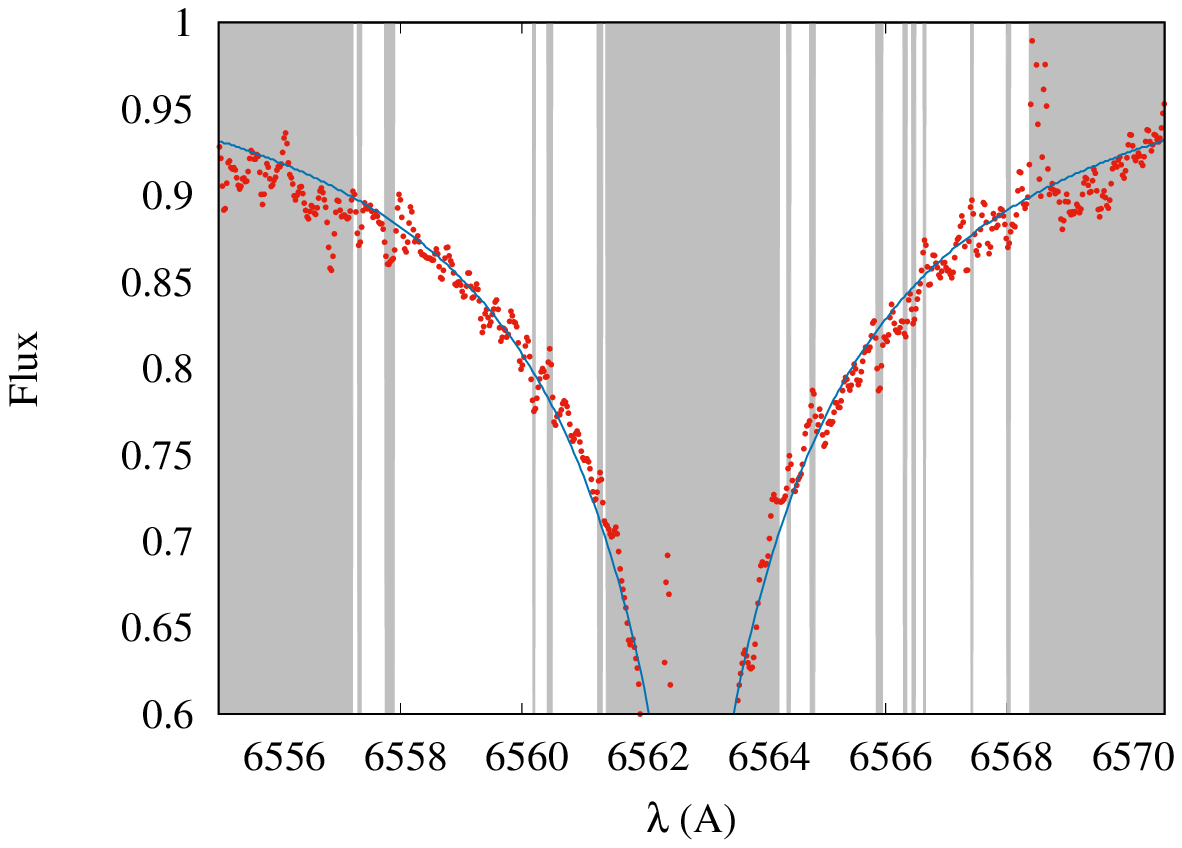}
\caption{Fitting results for the H$\beta$ line (left panel) and
H$\alpha$ line (right panel) of SDSS J1424+5615. Red points show the
observed normalized spectrum and the blue line shows the best-fit model
spectrum. To avoid the effects of cosmic rays and/or absorption lines, 
and to perform the fitting of the line wings,
the grey shaded regions are excluded from the fitting procedure 
(see text for details).\label{fighab}}
\end{figure*}

We finally check the fitting results by eye. The fitting results for
SDSS J1424+5615 are shown in Figure \ref{fighab}.

Although the microturbulent velocity ($v_t$) is not required for the
Balmer-line analysis, it needs to be determined for the abundance
analysis. The microturbulent velocity is not derived from the
Balmer-line profiles, but is determined so that abundances derived from
individual neutral Fe lines exhibit no trends with the strengths of the
lines. The uncertainty of $v_t$, expressed as $\delta v_t$, is
determined so that the trend is not significant at greater than the
$1\sigma$ level. In addition, we also examine the uncertainties of $v_t$
due to the errors in $T_{\rm eff}$ and $\log g$.

\subsubsection{Fe-Lines Method\label{sdssdiff}}

This method determines stellar parameters from an analysis of Fe absorption
lines in a spectrum, those that result in no dependence on the ionization stage,
excitation potential, or strength of the individual lines.

Suppose that $EW_i$ is the equivalent width of an Fe line and $A_i$ is
the Fe abundance determined from the line. In order to determine stellar
parameters, we evaluate three probabilities:
\begin{itemize}
\item[1] $p_{\rm ex}$: Probability that the correlation between completely uncorrelated 
sets of values becomes larger than the observed correlation between $A_i$ 
and excitation energy.
\item[2] $p_{\rm ion}$: Probability that a difference between Fe abundances determined
from neutral species and ionized species becomes larger than the observed difference
due only to measurement errors.
\item[3] $p_{\rm ew}$: Probability that a correlation between completely uncorrelated
set of values becomes larger than the observed correlation between $A_i$
and the normalized
equivalent width, $\log (EW_i/\lambda)$.
\end{itemize}
The probabilities $p_{\rm ex}$ and $p_{\rm ew}$ are evaluated using
Spearman's rank correlation test. This is because the expected
correlations are not necessarily linear, especially for the correlation
between $\log(EW_i/\lambda_i)$ and $A_i$. Then, we search for the
combination of stellar parameters ($T_{\rm eff},\,\log g,\,v_{\rm
turb}$) which maximizes $p=p_{\rm ex}\cdot p_{\rm ion}\cdot p_{\rm ew}$.
The $[\mathrm{Fe/H}]$ of the model atmosphere is also forced to agree with
the derived $[\mathrm{Fe/H}]$ within $0.3\,\mathrm{dex}$.

Uncertainties are estimated using a confidence-region boundary, where
$p=0.317p_{\max}$, assuming it to be a 3-D ellipsoid.
Covariances of any pair of the two parameters are also estimated.

Determination of stellar parameters from Fe lines fully relies on model
atmospheres, and could be significantly affected by deviation from local
thermodynamic equilibrium (non-LTE, NLTE) and the effect of 3D
motions in the atmosphere. The NLTE/3D effects might be significantly large
for EMP stars \citep{Asplund2005}. In a limited range
of $T_{\rm eff}$ and $[\mathrm{Fe/H}]$, however, the correction should
be systematic. 
For example, the difference between the Fe abundance derived 
from an NLTE analysis of the Fe I lines and that from an LTE analysis
varies less than $0.1\,\mathrm{dex}$ within the parameter range of 
our targets \citep{Lind2012}.
In order to avoid such systematic effects, we carry out
a line-by-line differential analysis adopting a well-studied bright
EMP turn-off star, G~64--12, as a reference star. For
each line, we first determine the difference in abundance ($\Delta A_i$)
between the target and G~64--12, and then convert it to the abundance of
the star $A_i$ by 
\begin{equation}
A_i = \langle A[\text{G 64--12}]\rangle + \Delta A_i.
\end{equation}

\subsubsection{SSPP-, Color-, \& Parallax-Based Stellar Parameter
Estimates}

Effective temperatures estimated by the SSPP, and given in Data Release
7 (DR7) of the SDSS, were adopted by \citet{Aoki2013} for their sample
of stars, from which our targets are selected. However, as the SSPP has
continued to be updated, here we adopt $T_{\rm eff}$ and $\log g$
estimates derived by the latest version.
The update results in higher $T_{\rm eff}$ by $\sim 100\,\mathrm{K}$.

We also derive effective temperatures from photometric colors, for a
comparison of stellar parameters estimated by different methods. We
first convert $g-,\,r-,\,{\rm and}\,i$-band \texttt{psfMag} measured in
the SDSS survey to the Johnson-Cousin $B,\,V,\,R_c,\,\text{and }I_c$ system using
the formulae provided by \citet{Jordi2006} for Population II stars.
Since SDSS photometry of G~64--12 and LP~815--43 suffer from saturation,
we adopt the APASS $V$ magnitudes for these two targets
\citep{Henden2016}. Infrared photometric data are taken from Two-Micron
All-Sky Survey \citep[2MASS; ][]{Cutri2003}. After correcting for
extinction from \citet{Schlafly2011}, we derive effective temperatures
from $V-K_s$ colors using the calibration of \citet{Casagrande2010},
with the assumption of $[\mathrm{Fe/H}]=-3.5$. These two methods are
based on calibrations using bright and/or nearby standard stars to
establish the scale. Since EMP turn-off stars are rare, the
uncertainties could be larger than those for less metal-poor stars.

The stars G~64--12 and LP~815--43 are both included in the Data Release 1 of Gaia.
We calculated their luminosities using the Gaia parallaxes, the bolometric
correction of \citet{Casagrande2010}, and $V$ and $K_s$ band magnitudes.
Surface gravity is then derived from the following equation:
\begin{equation}
\log g = \log g_{\odot} + \log (M/M_{\odot}) + 4\log (T_{\rm eff}/T_{\rm eff,\,\odot})-\log(L/L_{\odot}),
\end{equation}
where $M$ is the mass of the stars, assumed to be $0.75\,\mathrm{M_{\odot}}$, and $L$ is the luminosity of  the stars.
We adopt $\log g_{\odot} = 4.438$ and $T_{\rm eff} = 5777\,\mathrm{K}$ as the solar values.

\subsection{Results}

\subsubsection{Results for Bright EMP Stars}

Before discussing our SDSS program stars, we present results for
G~64--12 and LP~815--43, as a check on both the absolute and relative
scales of the derived stellar parameters in this work. Note that LP~815--43 is
analysed only for the evaluation of the relative scale, and is not
included in the subsequent abundance analysis.

Stellar parameter estimates for G~64--12 and LP~815--43 are summarized
in Table \ref{tablp815}. They are also shown in Figure \ref{fighrsdss}
and Figure \ref{figdtsdss}. We note that the result for the Balmer-line
analysis of G~64--12 is slightly different from our previous work
\citep{Matsuno2017}, due to small changes in the algorithm.  The
difference is still within the quoted uncertainty.

Results from the Balmer-line profile analysis agree with those obtained
from the Fe lines analysis within the errors. On the other hand, there
appears to be a systematic difference between the Balmer-line $T_{\rm
eff}$ estimates and the color-based $T_{\rm eff}$ estimates, by
$100-150\,\mathrm{K}$ \citep{Aoki2006,Norris2013}. A high-precision
analysis of G~64--12 has been carried out in previous studies
\citep{Placco2016,Reggiani2016}, and obtained $T_{\rm eff}=6463\,\mathrm{K}$
\citep{Melendez2010} and $\log g=4.26$ \citep{Nissen2007}. While our
temperature estimate based on the Balmer-line analysis is lower than
theirs, our color-based temperature estimate is consistent. See
\citet{Matsuno2017} for the detailed comparison among derived effective
temperatures of G~64--12 in previous studies.

Since a differential analysis is conducted in this work, the differences in
the estimated parameters between a target star and the reference star
(G~64--12) are important. The symbol ``$\Delta$'' listed in Table \ref{tablp815} describes 
this difference for parameters derived for G~64--12 and LP~815--43. The effective temperature of
LP~815--43 determined by each method is slightly higher than that of
G~64--12, whereas the $\log g$ of LP~815--43 is slightly lower than that
of G~64--12. We conclude that relative differences of parameters are not
so affected by the choice of the methods, even though a small offset
exists between individual methods. 

\begin{deluxetable}{lrrrr}
  \tablecaption{Comparison of Stellar Parameters for G~64--12 and LP~815--43\label{tablp815}}
  \tablehead{
   \colhead{Method}   & \colhead{Parameter}  & \colhead{G~64--12} & \colhead{LP~815--43} & \dcolhead{\Delta\tablenotemark{a}}
   } 
 \startdata
  Balmer lines            & $T_{\rm eff}\,\mathrm{(K)}$          & 6285   & 6323     &    38    \\  
                          & $\sigma_{T_{\rm eff}}\,\mathrm{(K)}$ & 26     & 31       &    40    \\        
                          & $\log g$               & 4.30   & 4.21     & -0.09    \\  
                          & $\sigma_{\log g}$      & 0.15   & 0.17     &  0.23    \\
                          & $v_t\,\mathrm{(km\,s^{-1})}$                  & 1.32   & 1.62     &  0.30    \\  
                          & $\sigma_{v_t}\,\mathrm{(km\,s^{-1})}$         & 0.18   & 0.19     &  0.26    \\
  Fe lines\tablenotemark{b}    & $T_{\rm eff}\,\mathrm{(K)}$          & (6285)   & 6424     &   139    \\  
                          & $\sigma_{T_{\rm eff}}\,\mathrm{(K)}$ & \nodata   &   83     &    79    \\  
                          & $\log g$               & (4.30)   & 4.22     & -0.08    \\  
                          & $\sigma_{\log g}$      & \nodata   & 0.20     &  0.13    \\
                          & $v_t\,\mathrm{(km\,s^{-1})}$                  & (1.32)   & 1.54     &  0.22    \\
                          & $\sigma_{v_t}\,\mathrm{(km\,s^{-1})}$         & \nodata   & 0.17     &  0.09    \\
  Color ($V-K_s$)         & $T_{\rm eff}\,\mathrm{(K)}$          & 6434   & 6481     &    47    \\  
                          & $\sigma_{T_{\rm eff}}\,\mathrm{(K)}$ & 55     & 67       &    87    \\  
  Parallax                & $\log g$               & 4.23   & 4.15     & -0.08    \\        
                          & $\sigma_{\log g}$      & 0.16   & 0.13     &  0.21    \\  
  \enddata
\tablenotetext{a}{Errors in $\Delta$ for the stellar parameters are
quadratic sums of the errors for the individual stars, except for the
Fe-lines method, for which errors of LP~815--43 are the quadratic sums
of the errors of G~64--12 and the error in $\Delta$.}

\tablenotetext{b}{Spectroscopic parameters are determined using G~64--12
as a reference star, using the listed parameters for this star.}
\end{deluxetable}

\subsubsection{Results for the SDSS Sample}

\begin{figure}
\plotone{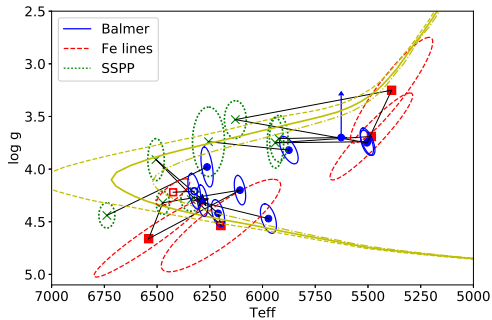}
\caption{Stellar parameters for our program stars in an H-R diagram,
together with $\alpha$-enhanced Y$^2$ isochrones with
$[\mathrm{Fe/H}]=-3.0$ and ages of $10,\,12,\,\mathrm{and}\,14\,\mathrm{Gyr}$, which are shown
as dot-dashed, solid, and dashed yellow lines, respectively
\citep{Kim2002}. The results of the analysis of Balmer-line profiles are
shown with blue filled circles, those of the Fe-lines analysis are shown
with red squares, and those of the SSPP are shown with green crosses.
The results for the same star are connected with black lines, and
$1\sigma$ uncertainties are shown with ellipses. The result of the
Balmer-line analysis for G~64--12 is shown with a blue star. LP~815--43
is shown with open symbols. The SDSS program stars are shown
with filled symbols.
\label{fighrsdss}}
\end{figure}

Results for our program sample of SDSS stars are summarized in Table
\ref{tabsdss} and shown in Figure \ref{fighrsdss}. Two out of the eight
stars contain only a small number of Fe absorption lines, and thus are
not suitable for spectroscopic determination of stellar parameters from
Fe lines. For two other stars, the set of stellar parameters which
simultaneously satisfy the three requirements listed in Section 
\ref{sdssdiff}
above are not found. The results based on the Fe-lines analysis for
these four stars are excluded from Table \ref{tabsdss}.

\begin{figure}
\plotone{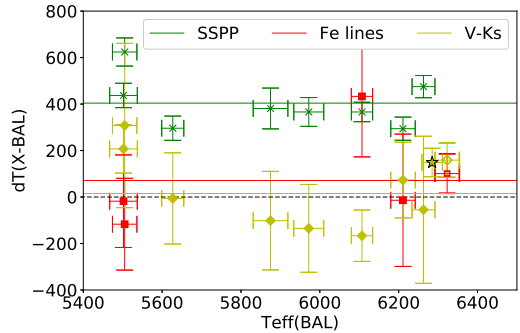}
\caption{Differences of the effective temperatures estimated by the analysis of
Balmer-line profiles from those obtained by other methods. The symbols
are the same as in Figure \ref{fighrsdss}, except yellow diamonds
indicate the results of the $V-K_s$ color-based estimate using the
approach of \citet{Casagrande2010}. Horizontal solid lines show the
unweighted mean of the differences for each method for the SDSS program
stars (for SSPP, $\langle \Delta T\rangle = 404\,\mathrm{K}$; for the
Fe-lines analysis, $\langle \Delta T\rangle = 71\,\mathrm{K}$; for
$V-K_s$ color, $\langle \Delta T\rangle = 15\,\mathrm{K}$). The black
dashed line shows $\Delta T = 0$.\label{figdtsdss}}
\end{figure}

A comparison of the results obtained by the different methods we
consider is presented in Figure~\ref{fighrsdss} and Figure
\ref{figdtsdss}. As can be seen, in most cases, the parameters from the
Fe-lines analysis are in agreement with those from the Balmer-line
profiles within the uncertainties, though the errors are large. We note
that no offset is expected due to the differential analysis. On the
other hand, there is a systematic difference in $T_{\rm eff}$ between
the SSPP-based estimates and the Balmer-line analysis, which could be
related to the known difference between $T_{\rm eff}$ from Balmer-line analysis 
and photometric $T_{\rm eff}$ estimates \citep{Norris2013}. 
We note that \citet{Aoki2013} adopted $T_{\rm eff}$ estimated by a previous 
version of SSPP, which are systematically lower than the present one.
Hence, the difference between our results of the Balmer line analysis
and those of \citet{Aoki2013} is about $100\,\mathrm{K}$ smaller.

Contrary to what was found from the analysis of G~64--12 and LP~815--43,
we do not find any systematic difference in effective temperature
estimates between the Balmer-line analysis and $V-K_s$ color approach. Although
\citet{Aoki2013} reached the conclusion that the offset in $T_{\rm eff}$
between $V-K_s$ color and the SSPP is small for their sample with
$[\mathrm{Fe/H}]<-2.5$, the latest version of the SSPP appears to
over-estimate $T_{\rm eff}$ compared to the $V-K_s$ color technique for
EMP turn-off stars. There is also a difference between the dust map
adopted to correct for interstellar extinction between the present
analysis and that used by \citet{Aoki2013}. We note that the comparison
here only includes a small sample in the present study, and that the
$T_{\rm eff}$ derived from colors could be affected by large errors in
the $K_s$ band magnitude of 2MASS for these fainter stars.

Uncertainties in the Balmer-line analysis are dominated by the
continuum placement, which does not significantly depend on $S/N$ for our
sample stars. Therefore, we adopt the results derived from 
the analysis of Balmer-line profiles in the abundance analysis described
below.  The Balmer-line analysis failed to determine the $\log g$ for SDSS
J0120--1001, due to the limited range of the grid (for $T_{\rm
eff}\geq 5600\,\mathrm{K}$, only $\log g \geq 3.4$ is covered). In
addition, the $\log g$ sensitivity of H$\alpha$ lines dramatically decreases
at $\log g \lesssim 3.4$. Considering the similarity of the SSPP $\log g$
between SDSS J1036+1212, SDSS J1522+3055, and SDSS J0120--1001, we adopt
$\log g = 3.4 \pm 0.3$ for the abundance analysis of this star.

\citet{Behara2010} also analyzed our program star SDSS J1036+1212.
Their $T_{\rm eff}$ estimate is based on the H$\alpha$ wing, and the
$\log g$ estimate is based on ionization balance between Fe~I and
Fe~II. They obtained $T_{\rm eff} = 6000\,\mathrm{K}$ and $\log g=4.0$,
which are $\sim 500\,\mathrm{K}$ and $0.3\,\mathrm{dex}$ higher than our
results. The $T_{\rm eff}$ derived from the $V-K_s$ color is somewhat 
closer, but still $300\,\mathrm{K}$ cooler than their result.
The source of the difference is not clear, because their
methods and ours are similar.  
There is another discrepancy between the present study and \citet{Behara2010}. 
It concerns the Eu abundance, and is discussed in Section \ref{empcemp}.

The adopted microturbulent velocities are listed in Table \ref{tabvtsdss}.
For SDSS J2309+2308 and SDSS J2005--1045, $v_t$ is not well-constrained
due to the small number of Fe absorption lines in these spectra. We
adopt $v_t=1.0\pm 0.5\,\mathrm{km\,s^{-1}}$ for these targets.

\begin{deluxetable*}{lrrrrrrrrrrrrrrrrrr}
  \tablecaption{Stellar Parameters for the SDSS Sample \label{tabsdss}}
\tablehead{      & \multicolumn{5}{c}{Balmer} && \multicolumn{4}{c}{Fe lines \tablenotemark{a}} && \multicolumn{4}{c}{SSPP} && \multicolumn{2}{c}{$V-K_s$} \\\cline{2-6}\cline{8-11}\cline{13-16}\cline{18-19}
\colhead{Object} & \colhead{$T_{\rm eff}$}    & \colhead{$\sigma$}                                  & \colhead{}               & \colhead{$\log g$}                                                              & \colhead{$\sigma$}   & & \colhead{$T_{\rm eff}$} & \colhead{$\sigma$} & \colhead{$\log g$} & \colhead{$\sigma$}& & \colhead{$T_{\rm eff}$} & \colhead{$\sigma$} & \colhead{$\log g$} & \colhead{ $\sigma$}& & \colhead{$T_{\rm eff}$} & \colhead{$\sigma$} \\
                 & \colhead{(K)}              & \colhead{(K)}                                       & \colhead{}               & \colhead{(dex)}                                                                 & \colhead{(dex)}      & & \colhead{(K)}           & \colhead{(K)}      & \colhead{(dex)}    & \colhead{(dex)}   & & \colhead{(K)}           & \colhead{(K)}      & \colhead{(dex)}    & \colhead{ (dex)}   & & \colhead{(K)} & \colhead{(K)}
}
\startdata
SDSS J0120--1001 & 5627 & 28 & $<$ & 3.70 \tablenotemark{b} & \nodata &  & \nodata & \nodata & \nodata & \nodata &  & 5923 & 44  & 3.71 & 0.22 &  & 5621 & 194               \\
SDSS J1036+1212  & 5502 & 35 &     & 3.74 & 0.13    &  & 5484    & 196     & 3.69    & 0.41    &  & 5939 & 39  & 3.75 & 0.24 &  & 5709 & 98               \\
SDSS J1424+5615  & 6107 & 27 &     & 4.20 & 0.14    &  & 6540    & 259     & 4.66    & 0.37    &  & 6473 & 32  & 4.32 & 0.11 &  & 5940 & 107               \\
SDSS J1522+3055  & 5505 & 31 &     & 3.75 & 0.12    &  & 5388    & 195     & 3.25    & 0.50    &  & 6129 & 52  & 3.53 & 0.18 &  & 5813 & 352              \\
SDSS J1640+3709  & 6211 & 31 &     & 4.42 & 0.13    &  & 6197    & 283     & 4.54    & 0.44    &  & 6505 & 39  & 3.91 & 0.20 &  & 6284 & 160              \\
SDSS J2005--1045 & 6263 & 29 &     & 3.98 & 0.19    &  & \nodata & \nodata & \nodata & \nodata &  & 6738 & 38  & 4.44 & 0.12 &  & 6208 & 315               \\
SDSS J2309+2308  & 5875 & 44 &     & 3.82 & 0.13    &  & \nodata & \nodata & \nodata & \nodata &  & 6256 &  76 & 3.74 & 0.33 &  & 5773 & 207               \\
SDSS J2349+3832  & 5972 & 38 &     & 4.47 & 0.14    &  & \nodata & \nodata & \nodata & \nodata &  & 6338 & 49  & 4.29 & 0.11 &  & 5837 & 185               \\
\enddata
\tablenotetext{a}{We use the stellar parameters of G~64--12 derived from Balmer line profiles as a reference.} 
\tablenotetext{b}{$3.40$ is adopted in the abundance analysis.} 
\end{deluxetable*}

\begin{deluxetable}{lrrrr}
\tablecaption{Microturbulent Velocities and Metallicities of Program
Stars\label{tabvtsdss}}
\tablehead{
 \colhead{Object} &\dcolhead{v_t}& \dcolhead{\sigma_{v_t}}& \dcolhead{[\mathrm{Fe/H}]}&\dcolhead{\sigma_{[\mathrm{Fe/H}]}}\\
      &\dcolhead{\mathrm{km\,s^{-1}}}&\dcolhead{\mathrm{km\,s^{-1}}}&& 
}
\startdata
G~64--12         & 1.32 & 0.18 & -3.38 & 0.02 \\
SDSS J0120--1001 & 0.99 & 0.48 & -3.84 & 0.06\\
SDSS J1036+1212  & 0.54 & 0.36 & -3.62 & 0.06\\
SDSS J1424+5615  & 0.97 & 0.24 & -3.10 & 0.03\\
SDSS J1522+3055  & 0.33 & 0.32 & -3.94 & 0.05\\
SDSS J1640+3709  & 1.19 & 0.38 & -3.54 & 0.05\\
SDSS J2005--1045 & 1.00 & 0.53 & -3.86 & 0.13\\
SDSS J2309+2308  & 1.00 & 0.50 & -3.96 & 0.09\\
SDSS J2349+3832  & 0.80 & 0.54 & -3.73 & 0.07\\
\enddata
\end{deluxetable}

\section{Chemical Abundances\label{sdsssecchem}}
\subsection{Abundance Analysis\label{sdsssecabun}}

We use 1D LTE model atmospheres from the ATLAS NEWODF grid with
$[\mathrm{\alpha/Fe}] = +0.4$ \citep{Castelli2003} in the abundance
analysis. Abundances are determined differentially with respect to the reference star
G~64--12, as described in Section \ref{sdssdiff}.

The line list used in the present work is based on that of
\citet{Aoki2013}, updated to include recently published $gf$-values. We then
restrict the list to the lines identified in the spectrum of G~64--12.
Because of a slight difference in instrument setting, our spectrum of
G~64--12 does not cover wavelengths shorter than $4100\,\mathrm{\AA}$. In
order to efficiently utilize the spectra of the SDSS sample, we also
include Mn~I $4044\,\mathrm{\AA}$, Fe~I $4046\,\mathrm{\AA}$, Fe~I
$4064\,\mathrm{\AA}$, and Sr~II $4078\,\mathrm{\AA}$, whose equivalent
widths of G~64--12 are taken from \citet{Reggiani2016}.

The equivalent width of each absorption line is measured by Gaussian
fitting, and listed in Table \ref{tabew}. The continuum level is
estimated by comparing the spectrum of the targets with that of
G~64--12. The widths of the absorption lines are dominated by instrumental
broadening and macroturbulence, and exclude the possibility of rapid
rotation for our sample stars.

In addition, we include the line list for the CH $G$-band from
\citet{Masseron2014} and the line list of $^7$Li from \citet{Smith1998}.
We determine abundances from spectrum synthesis using $4222.8-4325.4\,
\mathrm{\AA}$ for the CH $G$-band and $6707.4-6708.2\,\mathrm{\AA}$ for the Li
I doublet.

\startlongtable
\begin{deluxetable*}{lrrrrrr}
\tablecaption{Measured Equivalent Widths \label{tabew}}
\tablehead{
\colhead{Object}&\colhead{Species}&\colhead{Wavelength}&\colhead{ExPot}&\colhead{log(gf)}&\colhead{EW} & \colhead{Reference}\\    
                &            & \dcolhead{(\mathrm{\AA})}& \colhead{(eV)} &         & \dcolhead{(\mathrm{m\AA})} &  }
\startdata
 G~64--12         & Na I  & 5889.951 & 0.000 & 0.101  &   32.8& 1\\
 G~64--12         & Na I  & 5895.924 & 0.000 & -0.197 &   19.0& 1\\
 G~64--12         & Mg I  & 4167.271 & 4.346 & -0.710 &    4.0& 2\\
 G~64--12         & Mg I  & 4702.991 & 4.330 & -0.380 &    9.4& 2\\
 G~64--12         & Mg I  & 5172.684 & 2.712 & -0.450 &   77.4& 3\\
\enddata
\tablecomments{
1: \citet{Morton1991}, 2: \citet{FroeseFischer1975}, 3: \citet{Aldenius2007}, 4: \citet{Wiese1980}, 5: \citet{Smith1981}, 6: \citet{Ivans2006}, 7: \citet{Lawler1989}, 8: \citet{Piskunov1995}, 9: \citet{Lawler2013}, 10: \citet{Grevesse1989}, 11: \citet{Wood2013}, 12: \citet{Pickering2001}, 13: \citet{Ryabchikova1994}, 14: \citet{Martin1988}, 15: \citet{Sobeck2007}, 16: \citet{OBrian1991}, 17: \citet{Fuhr1988}, 18: \citet{Bard1991}, 19: \citet{Melendez2009}, 20: \citet{Moity1983}, 21: \citet{Pinnington1995}, 22: \citet{Grevesse1981}, 23: \citet{Grevesse2015}}
\tablecomments{Table \ref{tabew} is published in its entirety in the
machine-readable version. A portion is shown here for guidance regarding
its form and content.}
\end{deluxetable*}

We adopt the mean of the abundances determined from individual lines for
each species. Uncertainties are determined as follows:
\begin{equation}
 \sigma(X) = \sqrt{\sigma_{\rm lines}^2/N + \sigma_{\rm atm}^2},
\end{equation}
where $\sigma_{\rm lines}$ is the standard deviation of abundances
determined from individual lines, and $N$ is the number of lines used in
the analysis. When $N < 3$, we take $\sigma_{\rm FeI}$ as
$\sigma_{\rm lines}$. The variable $\sigma_{\rm atm}$ is the uncertainty due to
uncertainties in the stellar parameter estimates expressed as:
\begin{equation}
 \sigma_{\rm atm}^2 = \sum_{i=1}^{4}(\frac{\partial \log\epsilon}{\partial X_i}\sigma_{X_i})^2 + \sum_{i\neq j} \frac{\partial \log\epsilon}{\partial X_i}\frac{\partial \log\epsilon}{\partial X_j}\sigma_{X_i X_j},
\end{equation} 
where $(X_1,\,X_2,\,X_3,\,X_4)=(T_{\rm eff},\,\log g,\,v_t,\,[\mathrm{Fe/H}])$. 

In cases where no absorption lines are detected for a specific element,
we place conservative $5\sigma$ upper limits on the equivalent widths,
as the upper limits on equivalent widths do not contradict with the
equivalent widths of lines that are detected in the spectrum. These
$5\sigma$ upper limits are also placed on Li and C abundances from
spectral synthesis.

Note that our analysis is carried out differentially on the scale of
G~64--12. Therefore, when one compares the abundances with other papers,
it is required to include the uncertainty in abundances of G~64--12. In
particular, NLTE could significantly affect some elements, and 3D
effects may play a major role on the strengths of the molecular lines.

\subsection{Results}

Results of the abundance analysis are listed in Table \ref{tabsdssres}
(metallicities are listed in Table \ref{tabvtsdss}). The results are
also displayed in Figure \ref{figsdssres}. Note that the metallicities
for all but one of our program stars (7 of 8) are $[\mathrm{Fe/H}]<-3.5$.

The stars G~64--12 and SDSS J1424+5615 have been previously analyzed in
\citet{Matsuno2017}. As the line list and some algorithms employed have been
updated, the derived abundances for these stars are not exactly the same. The
difference is $\lesssim 0.05 \,\mathrm{dex}$ for G~64--12, and $\lesssim
0.2\,\mathrm{dex}$ for SDSS J1424+5615 ($\sim 0.4\,\mathrm{dex}$ for
$[\mathrm{C/Fe}]$). The larger difference for SDSS J1424+5615 is because
we derive abundances differentially in this study. Note that we are able
to derive a C abundance estimate for G~64--12, which was not derived in
\citet{Matsuno2017}. Although the redder part of the fitting region in our
spectrum is significantly affected by bad columns in the CCD, the bluer part
turns out to have sufficiently high quality to derive a C abundance
($\lesssim 4324.5\,\mathrm{\AA}$). The carbon ([C/Fe] $= +0.92$) and
barium ([Ba/Fe] $= -0.07$) abundance ratios of G~64--12 are consistent
with its classification as a CEMP star with no enhancement of
neutron-capture elements \citep[CEMP-no; ][]{Beers2005}, if we adopt
$[\mathrm{C/Fe}] > +0.7$ as the CEMP criterion
\citep{Placco2016}.

Whereas $\log g$ estimates in most previous studies are determined from
ionization balance of Fe I and Fe II lines, here we adopt $\log g $
estimated from Balmer-line profiles. We detect Fe II lines for seven
objects, for all of which the difference in Fe abundances from Fe II
lines and from Fe I lines is consistent with the difference of G~64--12
($0.11\, \mathrm{dex}$) within $2\sigma$, and for five of which the
differences are within $1\sigma$ (Figure \ref{figsdssres}). This result
indicates that the abundance results of the present work does not
essentially change even if the $\log g$ is estimated from a Fe I/II
balance.

\subsubsection{Carbon\label{sdsssecresc}}

Two stars in our SDSS program sample exhibit significant carbon
enhancement ($[\mathrm{C/Fe}]> +1.0$). One is SDSS J1036+1212
($[\mathrm{C/Fe}] = +1.19$) with a high Ba abundance ($[\mathrm{Ba/Fe}]
= +1.68$). The metallicity of this star is $[\mathrm{Fe/H}]=-3.62$,
which makes this star one of the most metal-poor CEMP-$s$ stars
\citep[CEMP stars with $s$-process enhancements; ][]{Beers2005}. The Sr
abundance of this star is low, which is a characteristic feature of
CEMP-$s$ stars. Although the abundances of SDSS J1036+1212 derived in
this work and those derived by \citet{Behara2010} differ, due to
different choice of effective temperature, high C, Ba and low Sr
abundances are obtained by both studies. The other CEMP star is SDSS
J1424+5615, which has been studied in \citet{Matsuno2017}.
Interestingly, both of the stars show large Na enhancements (Table
\ref{tabsdssres}). 

There is another star, SDSS J2309+2308, which exhibits Na and Ba
excesses, although its Ba abundance relies on only one Ba II line at
$4554\,\mathrm{\AA}$. This object could also be a CEMP-$s$ star. The upper
limit on C ($[\mathrm{C/Fe}] < +2.0$) is insufficient to determine
whether this star is C-rich or not. The reported upper limits on C
abundance for most of our SDSS program stars are not sufficiently low to
identify them as C-normal stars. 
   
Although Eu abundances for the C-rich stars is important to identify
``CEMP-i'' stars \citep{Hampel2016}, the relatively high temperatures of our
SDSS program stars prevents determination of meaningful limits for
their Eu abundances based on our present data.

\subsubsection{Lithium}

The Li abundance we obtain for G~64--12 places it on the Spite Plateau,
similar to those reported by previous studies, confirming that our
analysis is consistent in the framework of 1D/LTE analysis. SDSS
J1424+5615, which has the highest metallicity among stars in our sample
($[\mathrm{Fe/H}]=-3.10$), has a comparable Li abundance as G~64--12. By
contrast, all stars with $[\mathrm{Fe/H}]<-3.5$ in our sample have
lithium abundances less than $A(\mathrm{Li}) = 2.0$. This extends the
previously found decreasing trend of lithium toward the lowest
metallicity to $[\mathrm{Fe/H}]\sim -4$ \citep{Bonifacio2007,Aoki2009,
Sbordone2010}. The Li abundances are compared with previous studies and
discussed in detail in Section \ref{sdsssecdis}. 

\begin{figure*}
 \plotone{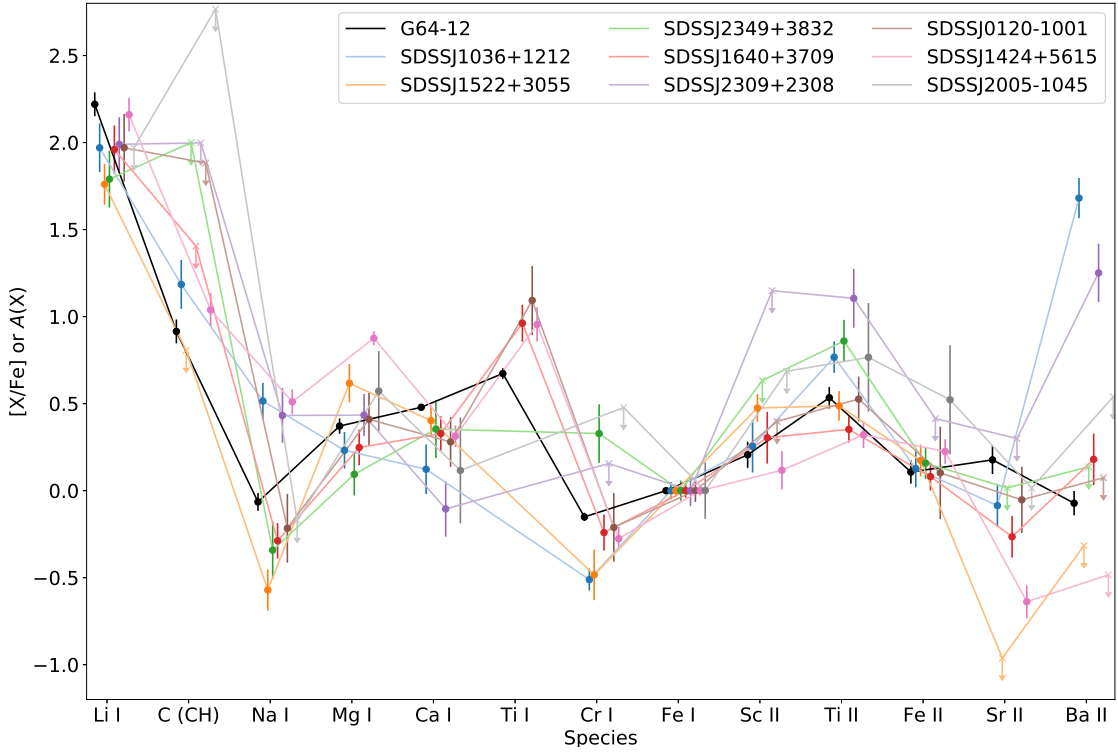}
 \caption{Chemical abundances of our program stars. Abundances are
determined differentially using G~64--12 as a reference (abundances are
shown with black dots and a solid line). The Li abundances are provided
as $A(\mathrm{Li})$, while the abundances of other elements are provided
as $[\mathrm{X/Fe}]$. Data points are slightly horizontally shifted for
clarity.\label{figsdssres}}
\end{figure*}

\startlongtable
\begin{deluxetable*}{llrrrrrrr}
  \tablecaption{Results of the Chemical Abundance Analysis \label{tabsdssres}}
   \tablehead{\colhead{Object} & \colhead{Species} & \colhead{$N$} &     & \colhead{$\log \epsilon (\mathrm{X})$} & \colhead{$\sigma$} &     & \colhead{$[\mathrm{X/Fe}]$\tablenotemark{a}} & \colhead{$\sigma$}}
   \startdata
   G~64--12                      & Li I              & 1             &     & 2.22                                   & 0.07              &     & 5.60               & 0.07\\
   G~64--12                      & C (CH)            & 1             &     & 5.97                                   & 0.07              &     & 0.92               & 0.07\\
   G~64--12                      & Na I              & 2             &     & 2.80                                   & 0.06              &     & -0.07              & 0.05\\
   G~64--12                      & Mg I              & 5             &     & 4.60                                   & 0.05              &     & 0.37               & 0.04\\
   G~64--12                      & Ca I              & 15            &     & 3.44                                   & 0.02              &     & 0.48               & 0.02\\
   G~64--12                      & Ti I              & 6             &     & 2.25                                   & 0.04              &     & 0.67               & 0.03\\
   G~64--12                      & Cr I              & 6             &     & 2.11                                   & 0.03              &     & -0.15              & 0.02\\
   G~64--12                      & Fe I              & 78            &     & 4.12                                   & 0.02              &     & 0.00               & 0.01\\
   G~64--12                      & Sc II             & 5             &     & -0.02                                  & 0.07              &     & 0.21               & 0.08\\
   G~64--12                      & Ti II             & 17            &     & 2.11                                   & 0.06              &     & 0.53               & 0.06\\
   G~64--12                      & Fe II             & 15            &     & 4.23                                   & 0.06              &     & 0.11               & 0.07\\
   G~64--12                      & Sr II             & 2             &     & -0.33                                  & 0.08              &     & 0.18               & 0.08\\
   G~64--12                      & Ba II             & 2             &     & -1.27                                  & 0.07              &     & -0.07              & 0.07\\\hline
   SDSS J0120--1001             & Li I              & 1             &     & 1.97                                   & 0.19              &     & 5.81               & 0.19\\
   SDSS J0120--1001             & C (CH)            &\nodata        & $<$ & 6.47                                   & \nodata             & $<$ & 1.88               & \nodata \\
   SDSS J0120--1001             & Na I              & 1             &     & 2.18                                   & 0.19              &     & -0.22              & 0.20\\
   SDSS J0120--1001             & Mg I              & 2             &     & 4.17                                   & 0.15              &     & 0.41               & 0.15\\
   SDSS J0120--1001             & Ca I              & 2             &     & 2.78                                   & 0.14              &     & 0.28               & 0.14\\
   SDSS J0120--1001             & Ti I              & 1             &     & 2.20                                   & 0.19              &     & 1.09               & 0.20\\
   SDSS J0120--1001             & Cr I              & 1             &     & 1.59                                   & 0.19              &     & -0.21              & 0.20\\
   SDSS J0120--1001             & Fe I              & 17            &     & 3.66                                   & 0.07              &     & 0.00               & 0.06\\
   SDSS J0120--1001             & Sc II             &\nodata        & $<$ & -0.29                                  & \nodata             & $<$ & 0.40               & \nodata \\
   SDSS J0120--1001             & Ti II             & 4             &     & 1.63                                   & 0.12              &     & 0.53               & 0.13\\
   SDSS J0120--1001             & Fe II             & 3             &     & 3.76                                   & 0.26              &     & 0.10               & 0.27\\
   SDSS J0120--1001             & Sr II             & 2             &     & -1.02                                  & 0.20              &     & -0.05              & 0.19\\
   SDSS J0120--1001             & Ba II             &\nodata        & $<$ & -1.59                                  & \nodata             & $<$ & 0.07               & \nodata \\\hline
   SDSS J1036+1212             & Li I              & 1             &     & 1.97                                   & 0.14              &     & 5.59               & 0.14\\
   SDSS J1036+1212             & C (CH)            & 1             &     & 6.00                                   & 0.14              &     & 1.19               & 0.14\\
   SDSS J1036+1212             & Na I              & 2             &     & 3.14                                   & 0.10              &     & 0.51               & 0.10\\
   SDSS J1036+1212             & Mg I              & 2             &     & 4.22                                   & 0.11              &     & 0.23               & 0.10\\
   SDSS J1036+1212             & Ca I              & 1             &     & 2.85                                   & 0.14              &     & 0.12               & 0.14\\
   SDSS J1036+1212             & Cr I              & 3             &     & 1.51                                   & 0.06              &     & -0.51              & 0.06\\
   SDSS J1036+1212             & Fe I              & 15            &     & 3.88                                   & 0.06              &     & 0.00               & 0.05\\
   SDSS J1036+1212             & Sc II             & 1             &     & -0.21                                  & 0.14              &     & 0.25               & 0.15\\
   SDSS J1036+1212             & Ti II             & 10            &     & 2.10                                   & 0.07              &     & 0.77               & 0.09\\
   SDSS J1036+1212             & Fe II             & 3             &     & 4.01                                   & 0.09              &     & 0.13               & 0.11\\
   SDSS J1036+1212             & Sr II             & 2             &     & -0.83                                  & 0.12              &     & -0.09              & 0.12\\
   SDSS J1036+1212             & Ba II             & 2             &     & 0.25                                   & 0.12              &     & 1.68               & 0.12\\\hline
   SDSS J1424+5615             & Li I              & 1             &     & 2.16                                   & 0.10              &     & 5.26               & 0.10\\
   SDSS J1424+5615             & C (CH)            & 1             &     & 6.37                                   & 0.10              &     & 1.04               & 0.10\\
   SDSS J1424+5615             & Na I              & 2             &     & 3.65                                   & 0.07              &     & 0.51               & 0.07\\
   SDSS J1424+5615             & Mg I              & 5             &     & 5.38                                   & 0.05              &     & 0.88               & 0.04\\
   SDSS J1424+5615             & Ca I              & 5             &     & 3.56                                   & 0.06              &     & 0.31               & 0.06\\
   SDSS J1424+5615             & Ti I              & 1             &     & 2.81                                   & 0.10              &     & 0.96               & 0.10\\
   SDSS J1424+5615             & Cr I              & 3             &     & 2.27                                   & 0.07              &     & -0.28              & 0.07\\
   SDSS J1424+5615             & Fe I              & 25            &     & 4.40                                   & 0.03              &     & 0.00               & 0.03\\
   SDSS J1424+5615             & Sc II             & 1             &     & 0.17                                   & 0.11              &     & 0.12               & 0.11\\
   SDSS J1424+5615             & Ti II             & 6             &     & 2.17                                   & 0.07              &     & 0.32               & 0.08\\
   SDSS J1424+5615             & Fe II             & 4             &     & 4.63                                   & 0.06              &     & 0.23               & 0.07\\
   SDSS J1424+5615             & Sr II             & 2             &     & -0.87                                  & 0.09              &     & -0.64              & 0.09\\
   SDSS J1424+5615             & Ba II             &\nodata        & $<$ & -1.40                                  & \nodata           & $<$ & -0.48              & \nodata \\\hline
   SDSS J1522+3055             & Li I              & 1             &     & 1.76                                   & 0.11              &     & 5.70               & 0.11\\
   SDSS J1522+3055             & C (CH)            &\nodata        & $<$ & 5.30                                   & \nodata             & $<$ & 0.81               & \nodata \\
   SDSS J1522+3055             & Na I              & 1             &     & 1.73                                   & 0.12              &     & -0.57              & 0.12\\
   SDSS J1522+3055             & Mg I              & 4             &     & 4.28                                   & 0.12              &     & 0.62               & 0.11\\
   SDSS J1522+3055             & Ca I              & 2             &     & 2.81                                   & 0.08              &     & 0.40               & 0.09\\
   SDSS J1522+3055             & Cr I              & 3             &     & 1.22                                   & 0.14              &     & -0.48              & 0.15\\
   SDSS J1522+3055             & Fe I              & 21            &     & 3.56                                   & 0.05              &     & 0.00               & 0.04\\
   SDSS J1522+3055             & Sc II             & 3             &     & -0.31                                  & 0.07              &     & 0.48               & 0.08\\
   SDSS J1522+3055             & Ti II             & 9             &     & 1.50                                   & 0.07              &     & 0.49               & 0.08\\
   SDSS J1522+3055             & Fe II             & 3             &     & 3.74                                   & 0.08              &     & 0.17               & 0.09\\
   SDSS J1522+3055             & Sr II             &\nodata        & $<$ & -2.03                                  & \nodata             & $<$ & -0.96              & \nodata \\
   SDSS J1522+3055             & Ba II             &\nodata        & $<$ & -2.07                                  & \nodata             & $<$ & -0.32              & \nodata \\\hline
   SDSS J1640+3709             & Li I              & 1             &     & 1.96                                   & 0.14              &     & 5.50               & 0.14\\
   SDSS J1640+3709             & C (CH)            &\nodata        & $<$ & 6.30                                   & \nodata             & $<$ & 1.41               & \nodata \\
   SDSS J1640+3709             & Na I              & 2             &     & 2.42                                   & 0.10              &     & -0.29              & 0.10\\
   SDSS J1640+3709             & Mg I              & 2             &     & 4.31                                   & 0.11              &     & 0.25               & 0.10\\
   SDSS J1640+3709             & Ca I              & 4             &     & 3.13                                   & 0.09              &     & 0.33               & 0.10\\
   SDSS J1640+3709             & Ti I              & 2             &     & 2.38                                   & 0.10              &     & 0.96               & 0.11\\
   SDSS J1640+3709             & Cr I              & 2             &     & 1.86                                   & 0.10              &     & -0.24              & 0.10\\
   SDSS J1640+3709             & Fe I              & 22            &     & 3.96                                   & 0.05              &     & 0.00               & 0.04\\
   SDSS J1640+3709             & Sc II             & 1             &     & -0.08                                  & 0.14              &     & 0.30               & 0.15\\
   SDSS J1640+3709             & Ti II             & 6             &     & 1.77                                   & 0.06              &     & 0.35               & 0.08\\
   SDSS J1640+3709             & Fe II             & 3             &     & 4.05                                   & 0.06              &     & 0.08               & 0.08\\
   SDSS J1640+3709             & Sr II             & 2             &     & -0.93                                  & 0.13              &     & -0.26              & 0.12\\
   SDSS J1640+3709             & Ba II             & 1             &     & -1.17                                  & 0.15              &     & 0.18               & 0.15\\\hline
   SDSS J2005--1045             & Li I              &\nodata        & $<$ & 1.97                                   & \nodata           & $<$ & 5.83               & \nodata \\
   SDSS J2005--1045             & C (CH)            &\nodata        & $<$ & 7.34                                   & \nodata           & $<$ & 2.77               & \nodata \\
   SDSS J2005--1045             & Na I              &\nodata        & $<$ & 2.21                                   & \nodata           & $<$ & -0.17              & \nodata \\
   SDSS J2005--1045             & Mg I              & 3             &     & 4.31                                   & 0.21              &     & 0.57               & 0.23\\
   SDSS J2005--1045             & Ca I              & 1             &     & 2.60                                   & 0.28              &     & 0.12               & 0.30\\
   SDSS J2005--1045             & Cr I              &\nodata        & $<$ & 2.26                                   & \nodata           & $<$ & 0.48               & \nodata \\
   SDSS J2005--1045             & Fe I              & 6             &     & 3.64                                   & 0.13              &     & 0.00               & 0.16\\
   SDSS J2005--1045             & Sc II             &\nodata        & $<$ & -0.02                                  & \nodata           & $<$ & 0.69               & \nodata \\
   SDSS J2005--1045             & Ti II             & 1             &     & 1.86                                   & 0.29              &     & 0.77               & 0.31\\
   SDSS J2005--1045             & Fe II             & 1             &     & 4.17                                   & 0.29              &     & 0.52               & 0.31\\
   SDSS J2005--1045             & Sr II             &\nodata        & $<$ & -0.97                                  & \nodata           & $<$ & 0.02               & \nodata \\
   SDSS J2005--1045             & Ba II             &\nodata        & $<$ & -1.14                                  & \nodata           & $<$ & 0.54               & \nodata \\\hline
   SDSS J2309+2308             & Li I              & 1             &     & 1.99                                   & 0.14              &     & 5.95               & 0.14\\
   SDSS J2309+2308             & C (CH)            &\nodata        & $<$ & 6.47                                   & \nodata             & $<$ & 2.00               & \nodata \\
   SDSS J2309+2308             & Na I              & 1             &     & 2.71                                   & 0.15              &     & 0.43               & 0.16\\
   SDSS J2309+2308             & Mg I              & 2             &     & 4.08                                   & 0.13              &     & 0.43               & 0.12\\
   SDSS J2309+2308             & Ca I              & 1             &     & 2.28                                   & 0.15              &     & -0.10              & 0.16\\
   SDSS J2309+2308             & Cr I              &\nodata        & $<$ & 1.84                                   & \nodata             & $<$ & 0.16               & \nodata \\
   SDSS J2309+2308             & Fe I              & 5             &     & 3.54                                   & 0.09              &     & 0.00               & 0.09\\
   SDSS J2309+2308             & Sc II             &\nodata        & $<$ & 0.34                                   & \nodata             & $<$ & 1.15               & \nodata \\
   SDSS J2309+2308             & Ti II             & 1             &     & 2.10                                   & 0.15              &     & 1.10               & 0.17\\
   SDSS J2309+2308             & Fe II             &\nodata        & $<$ & 3.96                                   & \nodata             & $<$ & 0.41               & \nodata \\
   SDSS J2309+2308             & Sr II             &\nodata        & $<$ & -0.79                                  & \nodata             & $<$ & 0.30               & \nodata \\
   SDSS J2309+2308             & Ba II             & 1             &     & -0.53                                  & 0.17              &     & 1.25               & 0.17\\\hline
   SDSS J2349+3832             & Li I              & 1             &     & 1.79                                   & 0.16              &     & 5.52               & 0.16\\
   SDSS J2349+3832             & C (CH)            &\nodata        & $<$ & 6.70                                   & \nodata             & $<$ & 2.00               & \nodata \\
   SDSS J2349+3832             & Na I              & 1             &     & 2.17                                   & 0.17              &     & -0.34              & 0.17\\
   SDSS J2349+3832             & Mg I              & 2             &     & 3.96                                   & 0.14              &     & 0.09               & 0.12\\
   SDSS J2349+3832             & Ca I              & 1             &     & 2.96                                   & 0.16              &     & 0.35               & 0.17\\
   SDSS J2349+3832             & Cr I              & 1             &     & 2.24                                   & 0.16              &     & 0.33               & 0.17\\
   SDSS J2349+3832             & Fe I              & 15            &     & 3.77                                   & 0.08              &     & 0.00               & 0.06\\
   SDSS J2349+3832             & Sc II             &\nodata        & $<$ & 0.05                                   & \nodata            & $<$ & 0.63               & \nodata \\
   SDSS J2349+3832             & Ti II             & 3             &     & 2.08                                   & 0.11              &     & 0.86               & 0.12\\
   SDSS J2349+3832             & Fe II             & 3             &     & 3.93                                   & 0.06              &     & 0.16               & 0.09\\
   SDSS J2349+3832             & Sr II             &\nodata        & $<$ & -0.84                                  & \nodata            & $<$ & 0.02               & \nodata \\
   SDSS J2349+3832             & Ba II             &\nodata        & $<$ & -1.41                                  & \nodata             & $<$ & 0.14               & \nodata \\\hline
\enddata
\tablecomments{Abundances are determined differentially using G~64--12 as a reference, and the uncertainties represent internal precision only.}
\tablenotetext{a}{$[\mathrm{X/Fe}]$ is calculated using the solar
abundance from \citet{Asplund2009}.}
\end{deluxetable*}

\section{Discussion\label{sdsssecdis}}
\subsection{Lithium Abundances of Extremely Metal-Poor stars}

\begin{figure*}
\plotone{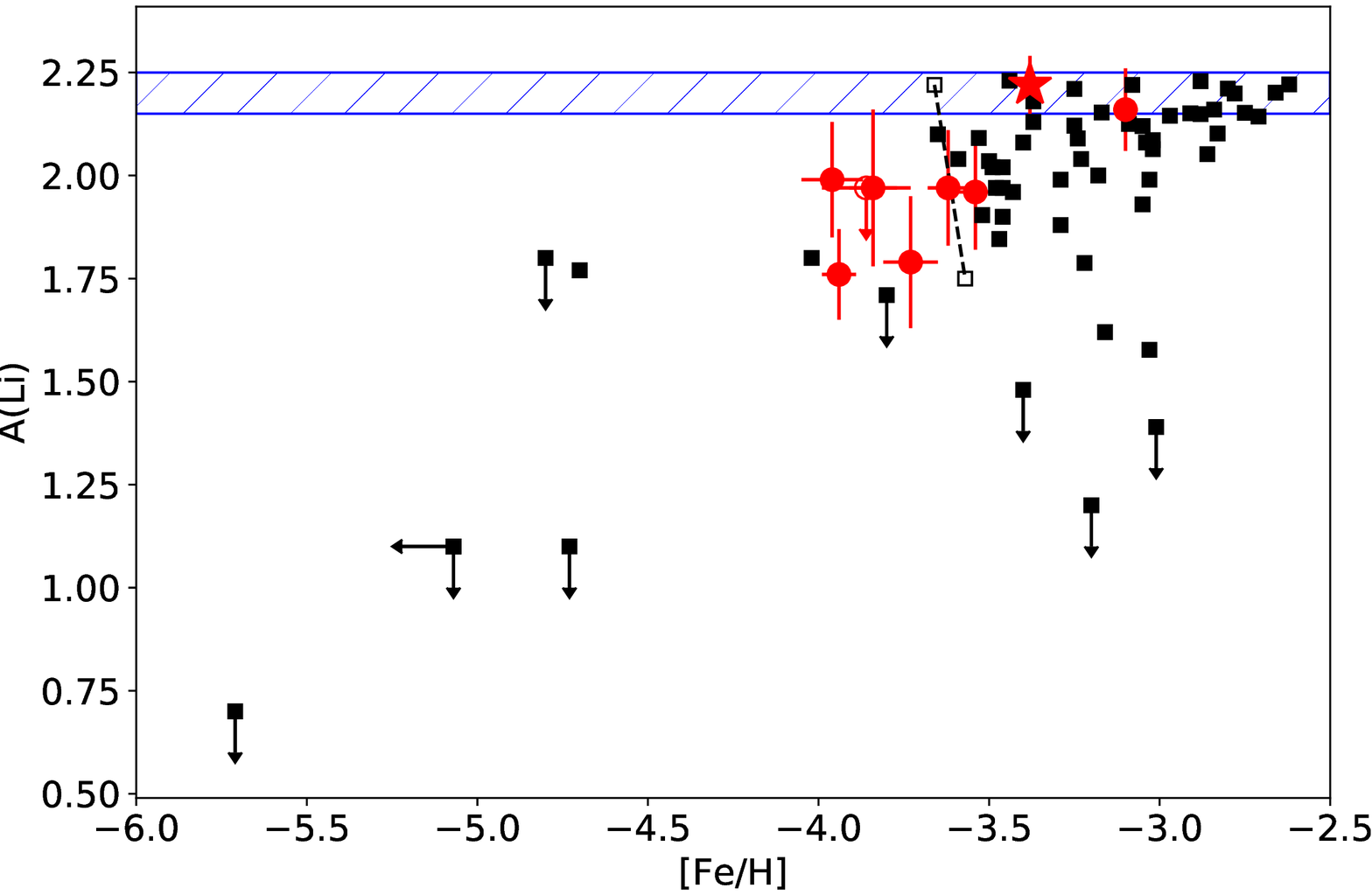}
\caption{$A(\mathrm{Li})$ as a function of $[\mathrm{Fe/H}]$. Our sample is shown by red 
circles for the SDSS program stars and a star for G~64--12. Literature data are
shown in black squares, compiled from \citet{Bonifacio2007,
Frebel2008, Aoki2009,Sbordone2010,Caffau2011,
Bonifacio2012,Bonifacio2015,Li2015a}. A double-lined spectroscopic binary system, CS 22876--032, is
shown by open squares with the individual Li abundances connected to
each other \citep{Norris2000,GonzalezHernandez2008}. 
The blue hatched region indicates the
Spite Plateau, $A(\mathrm{Li})=2.2\pm 0.1$.\label{figSDSSlife}}
\end{figure*}

\begin{figure*}
\plottwo{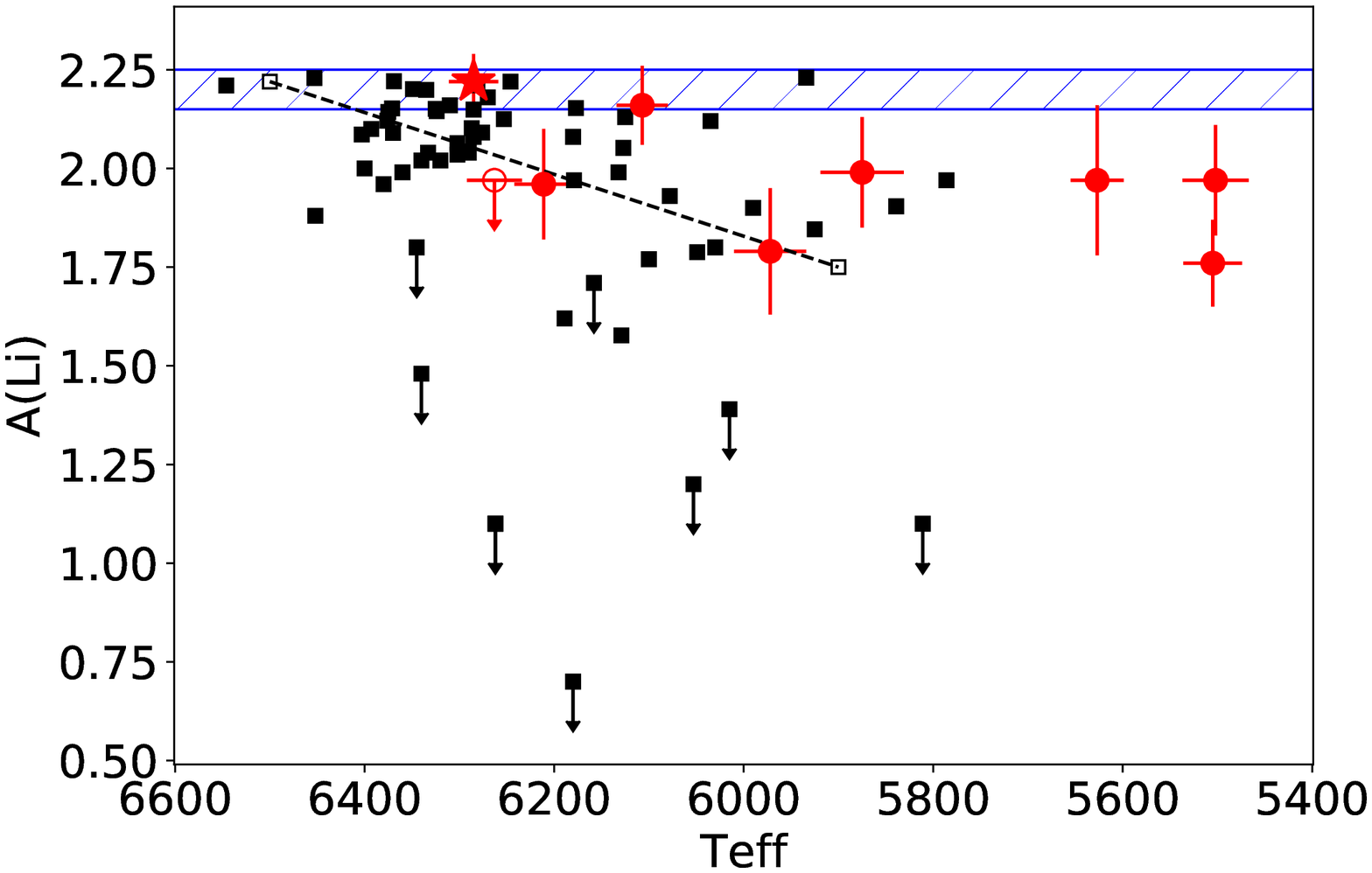}{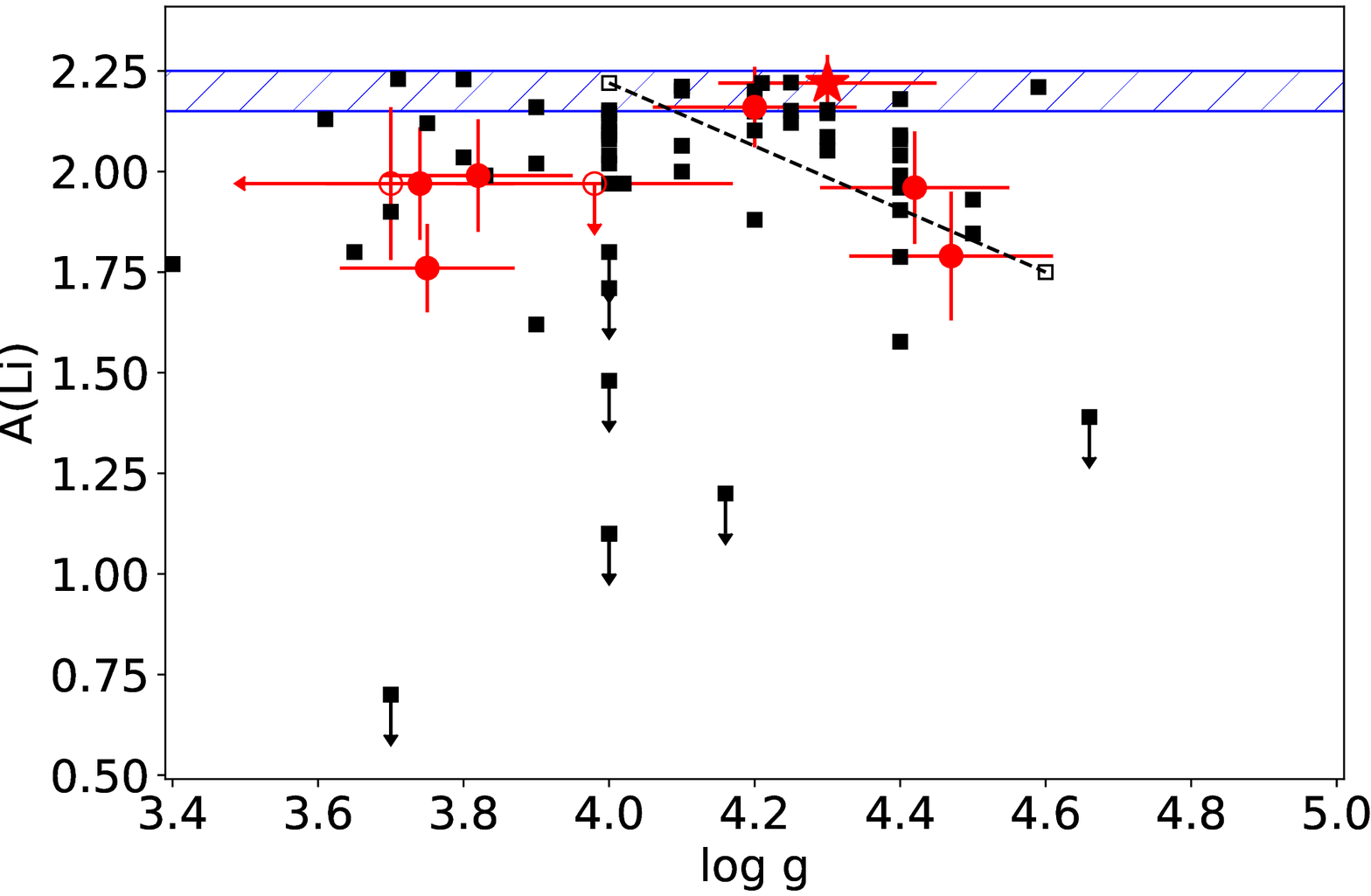}
\caption{$A(\mathrm{Li})$ as a function of $T_{\rm eff}$ and $\log g$. 
The symbols are the same as in Figure \ref{figSDSSlife}.\label{figSDSSlitg}}
\end{figure*}

The Li abundances of our program sample are shown in Figures \ref{figSDSSlife} and
\ref{figSDSSlitg}, together with stars from the literature. Previous Li
measurements and data selection for the plotting are summarized in the
Appendix. Our sample efficiently covers lower metallicities than most
previous samples, $[\mathrm{Fe/H}]<-3.5$. The average of the Li abundances below
$[\mathrm{Fe/H}]=-3.5$ in our sample is $\langle A(\mathrm{Li})\rangle =
1.90$, with a scatter of $0.10\,\mathrm{dex}$, which does not represent
a dispersion larger than can be accounted for by the errors of
determination.

One possible concern for interpretation of our results is the different
choice of temperature scale. Adopting $100\,\mathrm{K}$ hotter
temperatures makes the metallicities $0.08\,
\mathrm{dex}$ higher, and $A(\mathrm{Li})$ $0.08\,\mathrm{dex}$ higher.
A higher temperature scale by $\sim 300\,\mathrm{K}$ would bring the lithium
abundances of our sample onto the Spite Plateau level. However, our
analysis is carried out differentially to G~64--12, for which our analysis
of its Li abundance place it on the Spite Plateau. In addition, SDSS
J1424+5615, at $[\mathrm{Fe/H}]=-3.10$, was analysed by the same
procedure and has a Li abundance close to the plateau value. Considering
the Li abundances of G~64--12 and SDSS J1424+5615, the overall
temperature scale of our analysis is unlikely to be the reason for the
low Li abundances among the stars with lower metallicity.

We conclude that all stars in our sample with $[\mathrm{Fe/H}]<-3.5$
have lower Li abundance than the Spite Plateau, by $\sim 0.3\,
\mathrm{dex}$, with no scatter within the measurements errors. Hereafter
we combine our results with the literature sample.

As found from our sample, no star in the literature has comparable Li
abundance to the Spite Plateau below $[\mathrm{Fe/H}]\lesssim -3.5$,
except for the primary of the double-lined binary system CS 22876--032.
Thus, Li abundances appear to be uniformly low at extremely low
metallicity. Our sample fills in the gap between Li measurements for
stars around $[\mathrm{Fe/H}]\sim -3.5$ and the two previously studied
unevolved objects below $[\mathrm{Fe/H}]<-4.0$, LAMOST J1253+0753
\citep[$\mathrm{[Fe/H]}=-4.02$, $A(\mathrm{Li})=1.80$; ][]{Li2015a} and
HE 0233--0343 \citep[$\mathrm{[Fe/H]}=-4.7$, $A(\mathrm{Li})=1.77$;
][]{Hansen2014}.  The Li abundances are almost flat or slightly decreasing
from $[\mathrm{Fe/H}]\sim -3.5$ to $-4.5$, with relatively small scatter.
Below $[\mathrm{Fe/H}]\lesssim -5.0$, the observed Li abundances could show a sharp
drop, though there are only two stars that provide useful data.
Note that the possible existence of a ``dual plateau'' with a separation of
$\mathrm 0.1\,\mathrm{dex}$ are
discussed for stars with
$[\mathrm{Fe/H}]\gtrsim -2.5$ and $[\mathrm{Fe/H}]\lesssim -2.5$ by \citet{Melendez2010}.
The newly found constancy in the present work is at lower metallicity by $1.0\,\mathrm{dex}$ and with larger deviation from the Spite Plateau.

The combination of atomic diffusion and turbulent mixing has been
proposed as a means to explain the lower Li abundance for stars on the
Spite Plateau compared to the predicted primordial Li abundance
\citep{Richard2005,Michaud2015}. The model has been only tuned for
$[\mathrm{Fe/H}]=-2.3$ in \citet{Richard2005}. The key parameter is the
strength of the turbulent mixing, through which the model is adjusted.
Small differences in this parameter can cause a large surface Li
abundance difference \citep[see Fig. 3 in ][]{Richard2005}. If this
parameter has a metallicity dependence, this model might provide a
solution.

\citet{Fu2015} suggested that all Li in the atmospheres of metal-poor 
stars is accreted {\it after} star formation; they attributed the low
lithium abundances at $[\mathrm{Fe/H}]<-2.5$ to weak accretion. The
model needs fine tuning to reproduce the Spite Plateau, and it is not
yet clear whether it can account for our finding that all stars below
$[\mathrm{Fe/H}]=-3.5$ have lower Li abundances, with a small scatter.

\subsection{Chemical Inhomogeneity in the Early Universe}

The early Universe is expected to be chemically inhomogeneous, since
a small number of nucleosynthesis events can create
large abundance fluctuations from one place to another; the observed
scatter of the elemental abundances for EMP stars can be used to
quantify this inhomogeneity. The variations in yields from supernovae
explosions of the first stars \citep{Nomoto2013,Tominaga2014} are
primarily determined by differences in their explosion energies and
masses.

We have carried out $\chi^2$ tests to examine whether the observed
scatter in $[\mathrm{X/Fe}]$ (A(Li) for Li) is significant. The
probability that stars having the same abundance exhibit a scatter only due
to measurement errors is listed in
Table \ref{tabpvalsdss}. Stars without detection are excluded in this
analysis. We find statistically significant scatter for
$[\mathrm{Na/Fe}]$, $[\mathrm{Mg/Fe}]$, $[\mathrm{Cr/Fe}]$,
$[\mathrm{Ti/Fe}]$, $[\mathrm{Sr/Fe}]$, and $[\mathrm{Ba/Fe}]$. On the
other hand, $A(\mathrm{Li})$, $[\mathrm{Ca/Fe}]$ and $[\mathrm{Sc/Fe}]$
do not exhibit significant scatter even at extremely low metallicity,
although the number of stars with detection of Sc is small.
Thus, Ca and Fe seem to be produced at almost
a constant ratio irrespective of the progenitor. This ensures the
effectiveness of searches for metal-poor stars using Ca lines.
Since C is
detected only in three of our program stars, it is excluded from this
discussion.

The significant scatter observed for many elements indicate that the
natal clouds for early generation stars are chemically inhomogeneous,
reflecting variations in the yields of first stars and possible
variations in mixing. However, the scatter in these elements is small
compared to that predicted from the yields of supernovae explosions
, considering the mass range of the progenitor
\citep{Kobayashi2006}. The small scatter might indicate that the
mini-halos hosting the early formation of EMP stars are also polluted by
supernovae exploding in neighboring mini-halos 
\citep{Jeon2017}.

The abundance ratio $[\mathrm{Na/Fe}]$ apparently exhibits a bimodal distribution (Figure
\ref{figsdssres}). The Na abundances could be affected both by the
adopted analysis technique and internal processes intrinsic to a given
star. For instance, large NLTE effects in the
formation of Na I D lines have been predicted \citep{Andrievsky2007}.
However, the NLTE effect is almost systematic within the narrow
parameter range of our sample. Indeed, no significant difference in
stellar parameters is found between the stars with high and low Na
abundances. This stands in clear contrast to previous studies of EMP
stars, which often include red giants having a wide range of stellar
parameters compared to main-sequence turn-off stars. 
Another difficulty in the studies of red giants is that Na abundances 
could be affected by internal mixing during the evolution along the 
red giant branch.
Therefore, it has been difficult to
reach any conclusion about the Na abundance scatter from the sample
including red giants. Our study on turn-off stars provides a unique
sample to investigate the scatter and bimodal distribution of Na
abundance at the lowest metallicity. Hence, the bimodal distribution of
$[\mathrm{Na/Fe}]$ in our result is regarded not as a result of
analysis, but as a physical property of EMP stars.

In order to assess the origin of the observed bimodality, we examine possible
connections between the Na abundances and those of other elements. First,
the sample is divided into two groups at $[\mathrm{Na/Fe}]=0.0$. We
compute the probability that both sub-samples have the same mean
abundance, which is listed in the last column of Table
\ref{tabpvalsdss}. No significant difference is found for the abundances
within the measurement errors. Even if the same test is made excluding
SDSS J1424+5615, which has $>0.4\,\mathrm{dex}$ higher metallicity than
rest of the stars, the results remain the same.
Whereas correlations between $[\mathrm{Na/Fe}]$ and $A(\mathrm{Li})$ in
globular clusters has been reported \citep{Lind2009}, we find no
evidence for this in our sample.

Although the Na abundance may be related to a star's C abundance, as
discussed in Section \ref{sdsssecresc}, the difficulty in deriving C
abundances for the majority of our EMP turn-off stars prohibits a clear
conclusion. The $[\mathrm{Na/Fe}]$ bimodality and its association with C
abundance has been already reported \citep[e.g., ][]{Norris2013a}. Note,
however, that the difference in $[\mathrm{Na/Fe}]$ between their two
populations ($\Delta [\mathrm{Na/Fe}]\sim 2.0\,\mathrm{dex}$) is much
larger than ours ($\Delta [\mathrm{Na/Fe}]\sim 0.8\,
\mathrm{dex}$). The apparent $[\mathrm{Na/Fe}]$ bimodality should be confirmed and investigated
in detail by studies of a larger sample of EMP turn-off stars.

\begin{deluxetable}{lrrr}
\tablecaption{Scatter in $[\mathrm{X/Fe}]$ (or A(Li) for Li) and Probabilities\label{tabpvalsdss}}
\tablehead{
 \colhead{Element} &\colhead{Std. Deviation\tablenotemark{a}} & \dcolhead{p_{\rm Scatter}\tablenotemark{b}} & \dcolhead{p_{\rm Na-group}\tablenotemark{c}} 
}
\startdata
$[\mathrm{Fe/H}]$ & 0.33         & 0.00   & 0.14     \\ 
Li I              & 0.15         & 0.21   & 0.07     \\ 
Na I              & 0.48         & 0.00   & 0.00     \\ 
Mg I              & 0.30         & 0.00   & 0.21     \\ 
Ca I              & 0.15         & 0.16   & 0.19     \\ 
Cr I              & 0.26         & 0.00   & \nodata   \\ 
Sc II             & 0.16         & 0.06   & \nodata   \\ 
Ti II             & 0.26         & 0.00   & 0.65     \\ 
Sr II             & 0.28         & 0.00   & \nodata   \\ 
Ba II             & 0.79         & 0.00   & \nodata   \\ 
\enddata
\tablenotetext{a}{Stars without detection are excluded.}
\tablenotetext{b}{The probabilities that elemental abundances are the same for the whole sample (see Text).}
\tablenotetext{c}{The probabilities of Na-rich and Na-poor groups having the same abundances.}
\end{deluxetable}

\subsection{Metallicity Distribution Function}
\begin{figure}
\plotone{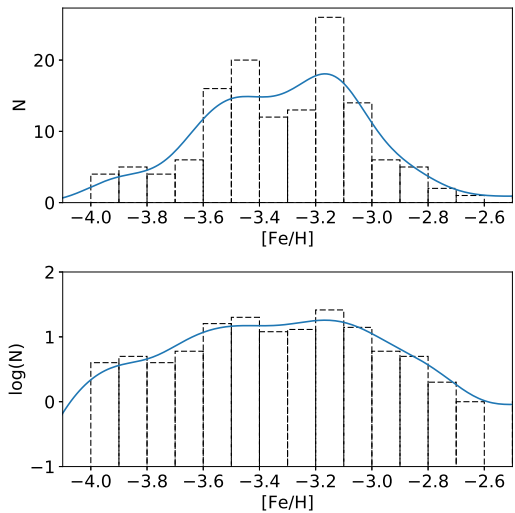}
\caption{The MDF of the full sample of SDSS main-sequence turnoff stars
from \citet{Aoki2013}, on a linear scale (upper panel) and on a logarithmic scale (lower panel). 
The solid blue lines are generalized histograms with a Gaussian convolution with $\sigma=0.10\,\mathrm{dex}$.
The metallicity of \citet{Aoki2013} is shifted by $0.28\,\mathrm{dex}$ to
match our metallicity scale.\label{SDSSMDF}}
\end{figure}

We now consider the metallicity distribution function (MDF) of the full sample of
\citet{Aoki2013}, shown in Figure \ref{SDSSMDF}. For the eight stars
re-analyzed in the present study, we replace the metallicities with the
newly derived ones. Since \citet{Aoki2013} adopt stellar parameters from
the SSPP, which derives higher $T_{\rm eff}$ than the present study,
there is a difference in the metallicity scale between the two results.
From a comparison of the stars in common between these studies
(excepting SDSS J2309+2308 and SDSS J2005--1045), our metallicity scale
is $\sim 0.28\,\mathrm{dex}$ lower and the $T_{\rm eff}$ scale is 
$\sim 300\,\mathrm{K}$ lower than that of \citet{Aoki2013}. Since
metallicity is lowered by $\sim 0.08\,\mathrm{dex}$ when $100\,
\mathrm{K}$ lower $T_{\rm eff}$ is adopted, the $0.28\,\mathrm{dex}$
offset is almost consistent with the value expected from our $\sim 350\,
\mathrm{K}$ cooler $T_{\rm eff}$ than the SSPP. 
We take a $0.28\,\mathrm{dex}$ shift into account in the replacement 
and generate the MDF
on the scale of the present study. The optimal bin size is determined
following \citet{Shimazaki2007}, $0.10\,\mathrm{dex}$. We also create
generalized histograms with a Gaussian function whose $\sigma$ is
$0.10\,\mathrm{dex}$. No significant spurious features are seen in the
distribution.
Note that our MDF above
$[\mathrm{Fe/H}]\sim -3.4$ appears to be significantly affected by the
incompleteness of the target selection \citep{Aoki2012} .

\citet{Schoerck2009} and \citet{Li2010} reported a cut-off in the MDFs
for giants and turn-off stars among candidate metal-poor stars
from the Hamburg/ESO survey\citep{Christlieb2008}, at around $[\mathrm{Fe/H}]\sim -3.5$. We
do not find evidence for such a cut-off in the MDF of our sample down to
$[\mathrm{Fe/H}]\sim -4.0$, consistent with \citet{Yong2013a}. If the
$0.28\,\mathrm{dex}$ metallicity correction is not applied, the
existence of the tail is still clear. 

\subsection{Extremely Metal-Poor CEMP-$s$ stars \label{empcemp}}

SDSS J1036+1212 is one of the lowest metallicity CEMP-$s$ stars known.
Although \citet{Behara2010} reported on a detailed abundance pattern for
SDSS J1036+1212, we could not detect as many elements as they reported.
We obtained the VLT/UVES spectrum of SDSS J1036+1212 used by
\citet{Behara2010} for their abundance analysis from the ESO archive.
However, we could not reproduce their reported detection of Eu. Hence,
we here discuss this object based solely on the abundance results
obtained by the present work.

SDSS J2309+2308 exhibits an excess of Ba, and is another candidate CEMP-$s$ star, 
although only a weak upper limit on its C abundance is determined by
our study.

CEMP-$s$ stars are generally considered to have experienced mass
transfer from an AGB companion in which large amounts of C and
$s$-process elements, including Ba, are synthesized. The reported high
binary frequency among such stars supports this scenario
\citep{Starkenburg2014,Hansen2016}. Although neither of these two stars
exhibited a radial velocity variation between our observations and
\citet{Aoki2013}, the radial velocity of SDSS J1036+1212 in our work is
$\sim 14\,\mathrm{km\,s^{-1}}$ larger than
\citet{Behara2010}, suggesting the likely binarity of this object.

\begin{figure*}
  \plottwo{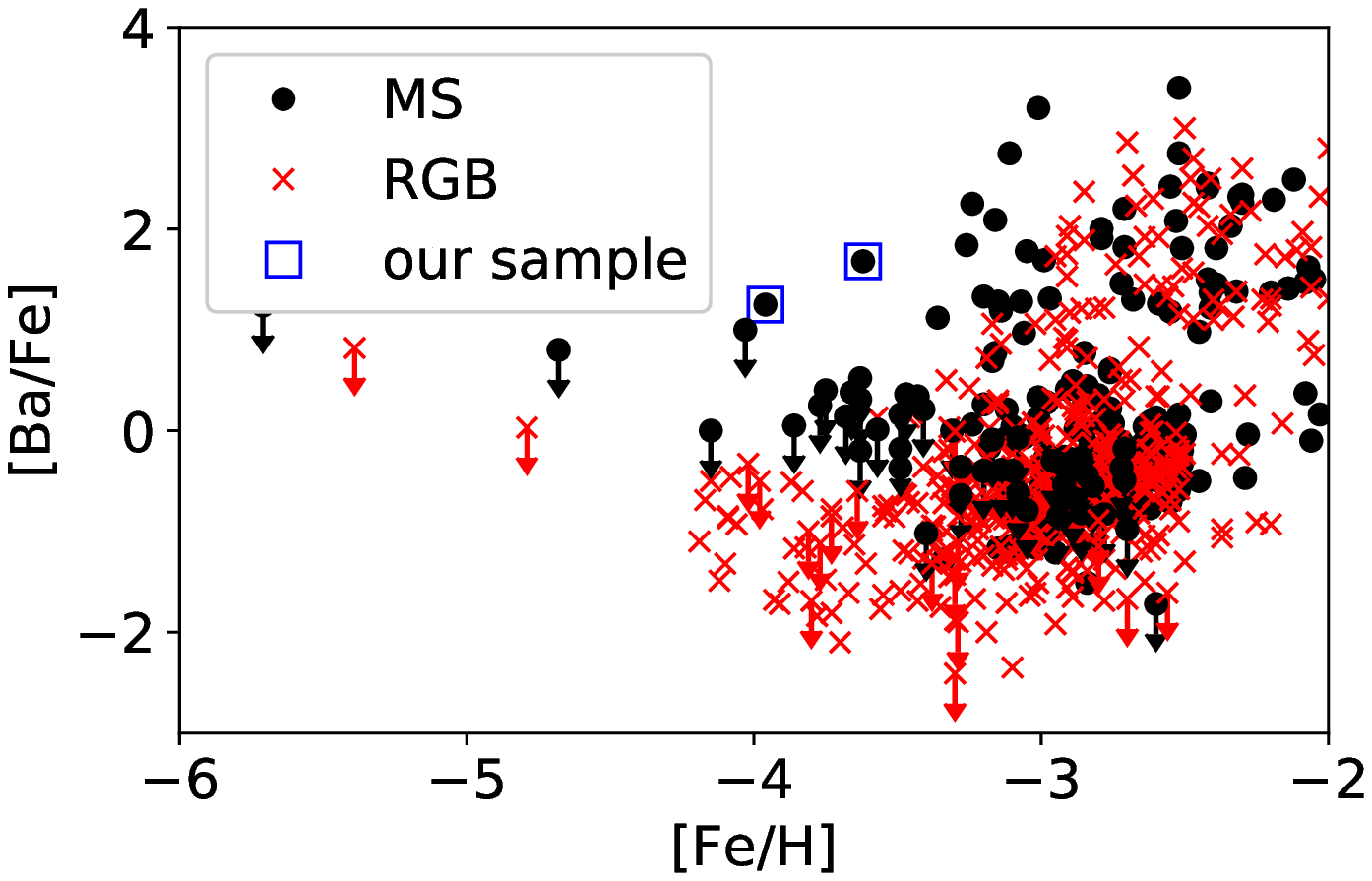}{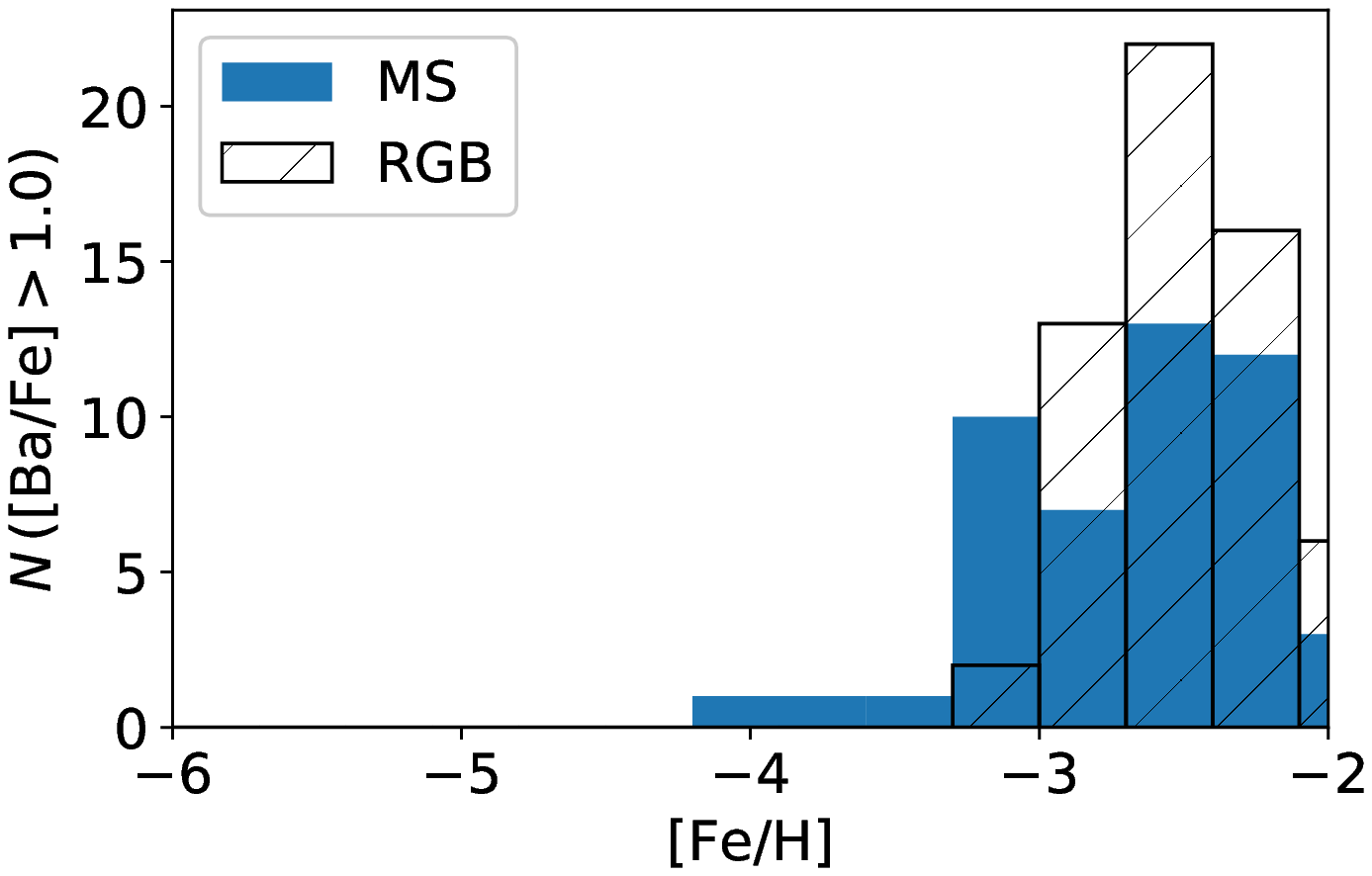}
  \caption{The distribution of [Ba/Fe], as a function of [Fe/H], for
stars selected from the SAGA database \citep{Suda2008} and the two
possible CEMP-$s$ stars in our sample. Classifications of main-sequence
stars (MS; black dots) and red giants (RGB; red crosses) are those
provided by SAGA. \textit{Left}: $[\mathrm{Ba/Fe}]$ as a function of
$[\mathrm{Fe/H}]$. \textit{Right}: Metallicity distribution of Ba-rich
stars ($[\mathrm{Ba/Fe}] > +1.0$). \label{SAGAba}}
\end{figure*}

Although two extremely metal-poor CEMP-$s$ candidates are found in our
sample, there is a lack of such stars among the red giants shown as found
Figure \ref{SAGAba}. Among the CEMP-$s$ stars with
$[\mathrm{Fe/H}]<-3.0$, almost all stars with $[\mathrm{Ba/Fe}] > +1.0$
are main-sequence stars. The lack of CEMP-$s$ red giants in
$[\mathrm{Fe/H}]<-3.0$ may be related to the first dredge-up that occurs
at the beginning of the red giant phase. First dredge-up dilutes the
surface material of a star to the
inner regions. If the over-abundance of Ba is provided by mass
transfer from a companion star to the stellar surface, first dredge-up
significantly reduces the surface Ba abundance. Such dilution is
effective only when i) the transferred mass is small compared to the
dredged-up mass, and ii) transferred material is not mixed with the
interior during main-sequence stage. Since the dredged-up mass is large
($\sim 50\,\%$ of the stellar mass), the first condition is generally
satisfied \citep[see also ][]{Masseron2012}. Therefore, the existence of
extremely metal-poor CEMP-$s$ turn-off stars is a potential constraint
on the efficiency of mixing processes during the main-sequence phase,
such as thermohaline mixing \citep{Stancliffe2007}. The reason for the
lack of CEMP-$s$ red giants at only extremely low metallicity might be
that the efficiency of the operation of the $s$-process is low in this
regime.  
  
Note that the above discussion using the SAGA database can be 
affected by sample selection, since
some past studies focus on red giants while others focus on turn-off stars.
For example, \citet{Jacobson2015} have constructed a sample of
metal-poor stars on the red giant branch using photometric estimates of
metallicity, and reported a lack of extremely carbon-rich objects.
\citet{Jacobson2015} suspected that the use of photometric selection of metal-poor stars 
may have resulted in a bias against such objects. On the other hand,
some previous studies focused on turn-off CEMP stars
\citep[e.g., ][]{Aoki2008}. To obtain a clear conclusion regarding the lack
of CEMP-$s$ red giants at $[\mathrm{Fe/H}] < -3$, we require a larger
sample of EMP stars including both red giants and turn-off stars.
 
\section{Summary\label{sdsssecsum}}

We analyze eight unevolved EMP stars for which
\citet{Aoki2013} have previously estimated abundances from snap-shot spectroscopy.
Based on newly obtained high-resolution, high-$S/N$ spectra, we first
compare different methods to derive stellar parameters. Analysis of
Balmer-line profiles derive consistent $T_{\rm eff}$ estimates with an Fe-lines
analysis and $V-K_s$ color-based temperature estimates. The surface
gravity estimates obtained from the Balmer-line analysis are also
consistent with those from the Fe-lines analysis and from Gaia
parallaxes. In contrast, the SSPP procedure results in higher $T_{\rm
eff}$ estimates than the Balmer-line analysis for EMP stars.

We carry out a differential abundance analysis, with G~64--12 as a
reference, adopting the parameters obtained by Balmer-line analysis. The
use of the reference star should cancel out NLTE/3D effects and atomic
data uncertainties. We obtain the following results:

\begin{enumerate}
\item Seven of the eight stars have $[\mathrm{Fe/H}]<-3.5$ and all have $T_{\rm eff}>5500\,\mathrm{K}$. 
\item Lithium abundances of all seven stars below $[\mathrm{Fe/H}] =
-3.5$ are lower than the Spite Plateau, without significant scatter.
This result could provide a constraint on proposed Li-depletion
mechanisms.
\item We found significant scatter in $[\mathrm{Na/Fe}]$, $[\mathrm{Mg/Fe}]$, $[\mathrm{Cr/Fe}]$, $[\mathrm{Ti/Fe}]$, $[\mathrm{Sr/Fe}]$ and $[\mathrm{Ba/Fe}]$. 
On the other hand, the scatter in $A(\mathrm{Li})$, $[\mathrm{Sc/Fe}]$,
and $[\mathrm{Ca/Fe}]$ are not significant. The observed bimodality in
$[\mathrm{Na/Fe}]$, with a separation of $0.8\,\mathrm{dex}$, requires
explanation; further confirmation and detailed investigation with larger
samples is desired.
\item We confirm the most metal-poor CEMP-$s$
star yet known and identify another CEMP-$s$ candidate with
$[\mathrm{Fe/H}]\sim -4.0$. 
From literature data, a lack of CEMP-$s$ red giants with
$[\mathrm{Fe/H}] < -3.0$ is seen. Their absence may be due to
the combined effect of metallicity dependence of $s$-process efficiency and dilution caused by first dredge-up.
\end{enumerate}

From the point of the comparison between observations and suggested Li-depletion
models, atomic diffusion with turbulent mixing \citep[e.g.,
][]{Richard2005} should be investigated for a wider range of parameter
space, especially toward lower metallicity. More quantitative evaluation
is needed for other models, such as Li depletion due to astration by first stars or Li
depletion in the pre-main sequence phase. More precise
stellar parameters, in particular evolutional phase and mass, and more
precise abundances are clearly desired. Improved $\log g$ estimates which
will be provided by Gaia parallaxes should result in significant progress.
It is also desired to increase the sample size of EMP
main-sequence turn-off stars. 

\acknowledgements

We thank H. Ito for the data reduction and observation. The authors are 
thankful to anonymous referee for the comments.
This work is 
based in part on data collected at Subaru Telescope and obtained from
the SMOKA, which is operated by the Astronomy Data Center, National
Astronomical Observatory of Japan. WA was supported by JSPS KAKENHI
Grant Numbers JP16H02168. This work received partial support from PHY
14--30152; Physics Frontier Center/JINA Center for the Evolution of the
Elements (JINA-CEE), awarded by the US National Science Foundation. This
work was supported in part by the Center for the Promotion of Integrated
Sciences (CPIS) of SOKENDAI. Y.S.L. acknowledges partial support from
the National Research Foundation of Korea to the Center for Galaxy
Evolution Research and Basic Science Research Program through the
National Research Foundation of Korea (NRF) funded by the Ministry of
Science, ICT and Future Planning (NRF-2015R1C1A1A02036658).

\appendix
\section{Adopted Values in the Tables}
Previous measurements of Li abundances for metal-poor stars are listed in Table \ref{tabliteli}.
We exclude stars with $T_{\rm eff}<5500\,\mathrm{K}$ or $[\mathrm{Fe/H}]>-2.5$.
There are overlaps in the samples among \citet{Aoki2009}, \citet{Sbordone2010} and \citet{Bonifacio2007}, for which we gave priority in this order.
Each star appears once in each figure, though we plot both of our
results and the results of \citet{Aoki2009} for G~64--12.
\startlongtable
\begin{deluxetable*}{lrrrrl}
 \tablecaption{Li Abundance Measurements from Previous Studies and in
this Work \label{tabliteli}}
 \tablehead{\colhead{Object}& \dcolhead{T_{\rm eff}}& \dcolhead{\log g} & \dcolhead{[\mathrm{Fe/H}]} & \dcolhead{A(\mathrm{Li})}& \colhead{Reference}}
\startdata
CD --24$^\circ$17504 & 6180 & 4.4   & -3.40    & 2.08    & \citet{Aoki2009}\\
BS 16023--046        & 6324 & 4.30  & -2.97    & 2.145   & \citet{Sbordone2010}\\
                    & 6364 & 4.50  & -2.97    & 2.18    & \citet{Bonifacio2007}\\
BS 16545--089        & 6320 & 3.9   & -3.49    & 2.02    & \citet{Aoki2009}\\
BS 16968--061        & 6035 & 3.75  & -3.05    & 2.12    & \citet{Bonifacio2007}\\
BS 17570--063        & 6078 & 4.50  & -3.05    & 1.930   & \citet{Sbordone2010}\\
                    & 6242 & 4.75  & -2.92    & 2.05    & \citet{Bonifacio2007}\\
BS 17572--100        & 6371 & 4.00  & -2.75    & 2.152   & \citet{Sbordone2010}\\
CS 22177--009        & 6177 & 4.30  & -3.17    & 2.153   & \citet{Sbordone2010}\\
                    & 6257 & 4.50  & -3.10    & 2.21    & \citet{Bonifacio2007}\\
CS 22188--033        & 6129 & 4.40  & -3.03    & 1.577   & \citet{Sbordone2010}\\
CS 22876--032A        & 6500 & 4.0   & -3.66    & 2.22    & \citet{GonzalezHernandez2008}\\
CS 22876--032B        & 5900 & 4.6   & -3.57    & 1.75    & \citet{GonzalezHernandez2008}\\
CS 22888--031        & 5925 & 4.50  & -3.47    & 1.846   & \citet{Sbordone2010}\\
                    & 6151 & 5.00  & -3.30    & 2.01    & \citet{Bonifacio2007}\\
CS 22948--093        & 6380 & 4.4   & -3.43    & 1.96    & \citet{Aoki2009}\\
                    & 6356 & 4.25  & -3.30    & 1.94    & \citet{Bonifacio2007}\\
                    & 6365 & 4.25  & -3.31    & 1.935   & \citet{Sbordone2010}\\
CS 22950--173        & 6335 & 4.20  & -2.78    & 2.199   & \citet{Sbordone2010}\\
CS 22953--037        & 6325 & 4.25  & -2.91    & 2.151   & \citet{Sbordone2010}\\
                    & 6364 & 4.25  & -2.89    & 2.16    & \citet{Bonifacio2007}\\
CS 22965--054        & 6310 & 3.9   & -2.84    & 2.16    & \citet{Aoki2009}\\
                    & 6089 & 3.75  & -3.04    & 2.03    & \citet{Bonifacio2007}\\
                    & 6245 & 4.00  & -2.90    & 2.161   & \citet{Sbordone2010}\\
CS 22966--011        & 6049 & 4.40  & -3.22    & 1.788   & \citet{Sbordone2010}\\
                    & 6204 & 4.75  & -3.07    & 1.90    & \citet{Bonifacio2007}\\
CS 29491--084        & 6285 & 4.00  & -3.04    & 2.080   & \citet{Sbordone2010}\\
                    & 6318 & 4.00  & -2.70    & 2.18    & \citet{Bonifacio2007}\\
CS 29499--060        & 6349 & 4.10  & -2.66    & 2.201   & \citet{Sbordone2010}\\
CS 29506--007        & 6285 & 4.20  & -2.88    & 2.149   & \citet{Sbordone2010}\\
                    & 6273 & 4.00  & -2.91    & 2.15    & \citet{Bonifacio2007}\\
CS 29506--090        & 6287 & 4.20  & -2.83    & 2.102   & \citet{Sbordone2010}\\
                    & 6303 & 4.25  & -2.83    & 2.12    & \citet{Bonifacio2007}\\
CS 29514--007        & 6281 & 4.10  & -2.80    & 2.211   & \citet{Sbordone2010}\\
CS 29516--028        & 5839 & 4.40  & -3.52    & 1.904   & \citet{Sbordone2010}\\
CS 29518--020        & 6127 & 4.30  & -2.86    & 2.052   & \citet{Sbordone2010}\\
                    & 6242 & 4.50  & -2.77    & 2.14    & \citet{Bonifacio2007}\\
CS 29518--043        & 6376 & 4.25  & -3.25    & 2.121   & \citet{Sbordone2010}\\
                    & 6432 & 4.25  & -3.20    & 2.17    & \citet{Bonifacio2007}\\
CS 29527--015        & 6276 & 4.00  & -3.53    & 2.091   & \citet{Sbordone2010}\\
                    & 6242 & 4.00  & -3.55    & 2.07    & \citet{Bonifacio2007}\\
CS 30301--024        & 6375 & 4.00  & -2.71    & 2.143   & \citet{Sbordone2010}\\
                    & 6334 & 4.00  & -2.75    & 2.12    & \citet{Bonifacio2007}\\
CS 30302--145        & 6403 & 4.30  & -3.02    & 2.086   & \citet{Sbordone2010}\\
CS 30339--069        & 6253 & 4.00  & -3.09    & 2.125   & \citet{Sbordone2010}\\
                    & 6242 & 4.00  & -3.08    & 2.13    & \citet{Bonifacio2007}\\
CS 30344--070        & 6302 & 4.10  & -3.02    & 2.064   & \citet{Sbordone2010}\\
CS 31061--032        & 6369 & 4.25  & -2.62    & 2.221   & \citet{Sbordone2010}\\
                    & 6409 & 4.25  & -2.58    & 2.25    & \citet{Bonifacio2007}\\
G 64--12             & 6270 & 4.4   & -3.37    & 2.18    & \citet{Aoki2009}\\
G 64--12             & 6285 & 4.30  & -3.38    & 2.22    & This work\\
G 64--37             & 6290 & 4.4   & -3.23    & 2.04    & \citet{Aoki2009}\\
HE 0148--2611        & 6400 & 4.10  & -3.18    & 2.000   & \citet{Sbordone2010}\\
HE 0233--0343        & 6100 & 3.4   & -4.7     & 1.77    & \citet{Hansen2014}\\
HE 1148--0037        & 5990 & 3.7   & -3.46    & 1.90    & \citet{Aoki2009}\\
HE 1327--2326        & 6180 & 3.7   & -5.71    & $<$0.70 & \citet{Frebel2008}\\
HE 1413--1954        & 6302 & 3.80  & -3.50    & 2.035   & \citet{Sbordone2010}\\
LAMOST J1253+0753   & 6030 & 3.65  & -4.02    & 1.80    & \citet{Li2015a}\\
LP 815--43           & 6453 & 3.80  & -2.88    & 2.229   & \citet{Sbordone2010}\\
SDSS J002113--005005 & 6546 & 4.59  & -3.25    & 2.21    & \citet{Bonifacio2012}\\
SDSS J002749+140418 & 6125 & 3.61  & -3.37    & 2.13    & \citet{Bonifacio2012}\\
SDSS J0040+16        & 6360 & 4.4   & -3.29    & 1.99    & \citet{Aoki2009}\\
SDSS J0120--1001    & 5627 & $<$3.70 & -3.84  & 1.97    & This work\\
SDSS J0212+0137     & 6333 & 4.0   & -3.59    & 2.04    & \citet{Bonifacio2015}\\
SDSS J031745+002304 & 5786 & 4.02  & -3.46    & 1.97    & \citet{Bonifacio2012}\\
SDSS J082118+181931 & 6158 & 4.00  & -3.80    & $<$1.71 & \citet{Bonifacio2012}\\
SDSS J082521+040334 & 6340 & 4.00  & -3.46    & 2.02    & \citet{Bonifacio2012}\\
SDSS J090733+024608 & 5934 & 3.71  & -3.44    & 2.23    & \citet{Bonifacio2012}\\
SDSS J102915+172927 & 5811 & 4.0   & -4.73    & $<$1.1  & \citet{Caffau2011}\\
SDSS J1033+40        & 6370 & 4.4   & -3.24    & 2.09    & \citet{Aoki2009}\\
SDSS J1035+0641     & 6262 & 4.0   & $<$-5.07 & $<$1.1  & \citet{Bonifacio2015}\\
SDSS J1036+1212     & 5502 & 3.74  & -3.62    & 1.97    & This work\\
SDSS J113528+010848 & 6132 & 3.83  & -3.03    & 1.99    & \citet{Bonifacio2012}\\
SDSS J1137+2553     & 6310 & 3.2   & -2.70    & 2.26    & \citet{Bonifacio2015}\\
SDSS J122935+262445 & 6452 & 4.20  & -3.29    & 1.88    & \citet{Bonifacio2012}\\
SDSS J130017+263238 & 6393 & 4.00  & -3.65    & 2.10    & \citet{Bonifacio2012}\\
SDSS J1424+5615     & 6107 & 4.20  & -3.10    & 2.16    & This work\\
SDSS J143632+091831 & 6340 & 4.00  & -3.40    & $<$1.48 & \citet{Bonifacio2012}\\
SDSS J144640+124917 & 6189 & 3.90  & -3.16    & 1.62    & \citet{Bonifacio2012}\\
SDSS J1522+3055     & 5505 & 3.75  & -3.94    & 1.76    & This work\\
SDSS J154246+054426 & 6179 & 4.00  & -3.48    & 1.97    & \citet{Bonifacio2012}\\
SDSS J1640+3709     & 6211 & 4.42  & -3.54    & 1.96    & This work\\
SDSS J1742+2531     & 6345 & 4.0   & -4.80    & $<$1.8  & \citet{Bonifacio2015}\\
SDSS J2005--1045    & 6263 & 3.98  & -3.86    & $<$1.97 & This work\\
SDSS J223143--094834 & 6053 & 4.16  & -3.20    & $<$1.20 & \citet{Bonifacio2012}\\
SDSS J230814--085526 & 6015 & 4.66  & -3.01    & $<$1.39 & \citet{Bonifacio2012}\\
SDSS J2309+2308      & 5875 & 3.82  & -3.96    & 1.99    & This work\\
SDSS J233113--010933 & 6246 & 4.21  & -3.08    & 2.22    & \citet{Bonifacio2012}\\
SDSS J2349+3832      & 5972 & 4.47  & -3.73    & 1.79    & This work\\
\enddata
\tablecomments{The first reference for each star in this table is
adopted in the plots (Figures \ref{figSDSSlife}, \ref{figSDSSlitg}). }
\end{deluxetable*}


\begin{thebibliography}{}
\expandafter\ifx\csname natexlab\endcsname\relax\def\natexlab#1{#1}\fi
\providecommand{\url}[1]{\href{#1}{#1}}

\bibitem[{{Aldenius} {et~al.}(2007){Aldenius}, {Tanner}, {Johansson},
  {Lundberg}, \& {Ryan}}]{Aldenius2007}
{Aldenius}, M., {Tanner}, J.~D., {Johansson}, S., {Lundberg}, H., \& {Ryan},
  S.~G. 2007, \aap, 461, 767

\bibitem[{{Allende Prieto} {et~al.}(2008){Allende Prieto}, {Sivarani}, {Beers},
  {Lee}, {Koesterke}, {Shetrone}, {Sneden}, {Lambert}, {Wilhelm}, {Rockosi},
  {Lai}, {Yanny}, {Ivans}, {Johnson}, {Aoki}, {Bailer-Jones}, \& {Re
  Fiorentin}}]{AllendePrieto2008}
{Allende Prieto}, C., {Sivarani}, T., {Beers}, T.~C., {et~al.} 2008, \aj, 136,
  2070

\bibitem[{{Andrievsky} {et~al.}(2007){Andrievsky}, {Spite}, {Korotin}, {Spite},
  {Bonifacio}, {Cayrel}, {Hill}, \& {Fran{\c c}ois}}]{Andrievsky2007}
{Andrievsky}, S.~M., {Spite}, M., {Korotin}, S.~A., {et~al.} 2007, \aap, 464,
  1081

\bibitem[{{Aoki}(2012)}]{Aoki2012}
{Aoki}, W. 2012, in Astronomical Society of the Pacific Conference Series, Vol.
  458, Galactic Archaeology: Near-Field Cosmology and the Formation of the
  Milky Way, ed. W.~{Aoki}, M.~{Ishigaki}, T.~{Suda}, T.~{Tsujimoto}, \&
  N.~{Arimoto}, 55

\bibitem[{{Aoki} {et~al.}(2009){Aoki}, {Barklem}, {Beers}, {Christlieb},
  {Inoue}, {Garc{\'{\i}}a P{\'e}rez}, {Norris}, \& {Carollo}}]{Aoki2009}
{Aoki}, W., {Barklem}, P.~S., {Beers}, T.~C., {et~al.} 2009, \apj, 698, 1803

\bibitem[{{Aoki} {et~al.}(2006){Aoki}, {Frebel}, {Christlieb}, {Norris},
  {Beers}, {Minezaki}, {Barklem}, {Honda}, {Takada-Hidai}, {Asplund}, {Ryan},
  {Tsangarides}, {Eriksson}, {Steinhauer}, {Deliyannis}, {Nomoto}, {Fujimoto},
  {Ando}, {Yoshii}, \& {Kajino}}]{Aoki2006}
{Aoki}, W., {Frebel}, A., {Christlieb}, N., {et~al.} 2006, \apj, 639, 897

\bibitem[{{Aoki} {et~al.}(2008){Aoki}, {Beers}, {Sivarani}, {Marsteller},
  {Lee}, {Honda}, {Norris}, {Ryan}, \& {Carollo}}]{Aoki2008}
{Aoki}, W., {Beers}, T.~C., {Sivarani}, T., {et~al.} 2008, \apj, 678, 1351

\bibitem[{{Aoki} {et~al.}(2013){Aoki}, {Beers}, {Lee}, {Honda}, {Ito},
  {Takada-Hidai}, {Frebel}, {Suda}, {Fujimoto}, {Carollo}, \&
  {Sivarani}}]{Aoki2013}
{Aoki}, W., {Beers}, T.~C., {Lee}, Y.~S., {et~al.} 2013, \aj, 145, 13

\bibitem[{{Asplund}(2005)}]{Asplund2005}
{Asplund}, M. 2005, \araa, 43, 481

\bibitem[{{Asplund} {et~al.}(2009){Asplund}, {Grevesse}, {Sauval}, \&
  {Scott}}]{Asplund2009}
{Asplund}, M., {Grevesse}, N., {Sauval}, A.~J., \& {Scott}, P. 2009, \araa, 47,
  481

\bibitem[{{Baba} {et~al.}(2002){Baba}, {Yasuda}, {Ichikawa}, {Yagi}, {Iwamoto},
  {Takata}, {Horaguchi}, {Taga}, {Watanabe}, {Ozawa}, \& {Hamabe}}]{Baba2002}
{Baba}, H., {Yasuda}, N., {Ichikawa}, S.-I., {et~al.} 2002, in Astronomical
  Society of the Pacific Conference Series, Vol. 281, Astronomical Data
  Analysis Software and Systems XI, ed. D.~A. {Bohlender}, D.~{Durand}, \&
  T.~H. {Handley}, 298

\bibitem[{{Bard} {et~al.}(1991){Bard}, {Kock}, \& {Kock}}]{Bard1991}
{Bard}, A., {Kock}, A., \& {Kock}, M. 1991, \aap, 248, 315

\bibitem[{{Barklem} {et~al.}(2002){Barklem}, {Stempels}, {Allende Prieto},
  {Kochukhov}, {Piskunov}, \& {O'Mara}}]{Barklem2002}
{Barklem}, P.~S., {Stempels}, H.~C., {Allende Prieto}, C., {et~al.} 2002, \aap,
  385, 951

\bibitem[{{Beers} \& {Christlieb}(2005)}]{Beers2005}
{Beers}, T.~C., \& {Christlieb}, N. 2005, \araa, 43, 531

\bibitem[{{Behara} {et~al.}(2010){Behara}, {Bonifacio}, {Ludwig}, {Sbordone},
  {Gonz{\'a}lez Hern{\'a}ndez}, \& {Caffau}}]{Behara2010}
{Behara}, N.~T., {Bonifacio}, P., {Ludwig}, H.-G., {et~al.} 2010, \aap, 513,
  A72

\bibitem[{{Bonifacio} {et~al.}(2012){Bonifacio}, {Sbordone}, {Caffau},
  {Ludwig}, {Spite}, {Gonz{\'a}lez Hern{\'a}ndez}, \& {Behara}}]{Bonifacio2012}
{Bonifacio}, P., {Sbordone}, L., {Caffau}, E., {et~al.} 2012, \aap, 542, A87

\bibitem[{{Bonifacio} {et~al.}(2007){Bonifacio}, {Molaro}, {Sivarani},
  {Cayrel}, {Spite}, {Spite}, {Plez}, {Andersen}, {Barbuy}, {Beers}, {Depagne},
  {Hill}, {Fran{\c c}ois}, {Nordstr{\"o}m}, \& {Primas}}]{Bonifacio2007}
{Bonifacio}, P., {Molaro}, P., {Sivarani}, T., {et~al.} 2007, \aap, 462, 851

\bibitem[{{Bonifacio} {et~al.}(2015){Bonifacio}, {Caffau}, {Spite}, {Limongi},
  {Chieffi}, {Klessen}, {Fran{\c c}ois}, {Molaro}, {Ludwig}, {Zaggia}, {Spite},
  {Plez}, {Cayrel}, {Christlieb}, {Clark}, {Glover}, {Hammer}, {Koch},
  {Monaco}, {Sbordone}, \& {Steffen}}]{Bonifacio2015}
{Bonifacio}, P., {Caffau}, E., {Spite}, M., {et~al.} 2015, \aap, 579, A28

\bibitem[{{Caffau} {et~al.}(2011){Caffau}, {Bonifacio}, {Fran{\c c}ois},
  {Sbordone}, {Monaco}, {Spite}, {Spite}, {Ludwig}, {Cayrel}, {Zaggia},
  {Hammer}, {Randich}, {Molaro}, \& {Hill}}]{Caffau2011}
{Caffau}, E., {Bonifacio}, P., {Fran{\c c}ois}, P., {et~al.} 2011, \nat, 477,
  67

\bibitem[{{Casagrande} {et~al.}(2010){Casagrande}, {Ram{\'{\i}}rez},
  {Mel{\'e}ndez}, {Bessell}, \& {Asplund}}]{Casagrande2010}
{Casagrande}, L., {Ram{\'{\i}}rez}, I., {Mel{\'e}ndez}, J., {Bessell}, M., \&
  {Asplund}, M. 2010, \aap, 512, A54

\bibitem[{{Castelli} \& {Kurucz}(2003)}]{Castelli2003}
{Castelli}, F., \& {Kurucz}, R.~L. 2003, in IAU Symposium, Vol. 210, Modelling
  of Stellar Atmospheres, ed. N.~{Piskunov}, W.~W. {Weiss}, \& D.~F. {Gray},
  A20

\bibitem[{{Christlieb} {et~al.}(2008){Christlieb}, {Sch{\"o}rck}, {Frebel},
  {Beers}, {Wisotzki}, \& {Reimers}}]{Christlieb2008}
{Christlieb}, N., {Sch{\"o}rck}, T., {Frebel}, A., {et~al.} 2008, \aap, 484,
  721

\bibitem[{{Coc} {et~al.}(2004){Coc}, {Vangioni-Flam}, {Descouvemont},
  {Adahchour}, \& {Angulo}}]{Coc2004}
{Coc}, A., {Vangioni-Flam}, E., {Descouvemont}, P., {Adahchour}, A., \&
  {Angulo}, C. 2004, \apj, 600, 544

\bibitem[{{Cutri} {et~al.}(2003){Cutri}, {Skrutskie}, {van Dyk}, {Beichman},
  {Carpenter}, {Chester}, {Cambresy}, {Evans}, {Fowler}, {Gizis}, {Howard},
  {Huchra}, {Jarrett}, {Kopan}, {Kirkpatrick}, {Light}, {Marsh}, {McCallon},
  {Schneider}, {Stiening}, {Sykes}, {Weinberg}, {Wheaton}, {Wheelock}, \&
  {Zacarias}}]{Cutri2003}
{Cutri}, R.~M., {Skrutskie}, M.~F., {van Dyk}, S., {et~al.} 2003, VizieR Online
  Data Catalog, 2246

\bibitem[{{Cyburt} {et~al.}(2016){Cyburt}, {Fields}, {Olive}, \&
  {Yeh}}]{Cyburt2016}
{Cyburt}, R.~H., {Fields}, B.~D., {Olive}, K.~A., \& {Yeh}, T.-H. 2016, Reviews
  of Modern Physics, 88, 015004

\bibitem[{{Frebel} {et~al.}(2008){Frebel}, {Collet}, {Eriksson}, {Christlieb},
  \& {Aoki}}]{Frebel2008}
{Frebel}, A., {Collet}, R., {Eriksson}, K., {Christlieb}, N., \& {Aoki}, W.
  2008, \apj, 684, 588

\bibitem[{{Froese Fischer}(1975)}]{FroeseFischer1975}
{Froese Fischer}, C. 1975, Canadian Journal of Physics, 53, 184, (FFa)

\bibitem[{{Fu} {et~al.}(2015){Fu}, {Bressan}, {Molaro}, \& {Marigo}}]{Fu2015}
{Fu}, X., {Bressan}, A., {Molaro}, P., \& {Marigo}, P. 2015, \mnras, 452, 3256

\bibitem[{{Fuhr} {et~al.}(1988){Fuhr}, {Martin}, \& {Wiese}}]{Fuhr1988}
{Fuhr}, J.~R., {Martin}, G.~A., \& {Wiese}, W.~L. 1988, Journal of Physical and
  Chemical Reference Data, 17

\bibitem[{{Gaia Collaboration} {et~al.}(2016{\natexlab{a}}){Gaia
  Collaboration}, {Brown}, {Vallenari}, {Prusti}, {de Bruijne}, {Mignard},
  {Drimmel}, {Babusiaux}, {Bailer-Jones}, {Bastian}, \&
  et~al.}]{GaiaCollaboration2016}
{Gaia Collaboration}, {Brown}, A.~G.~A., {Vallenari}, A., {et~al.}
  2016{\natexlab{a}}, \aap, 595, A2

\bibitem[{{Gaia Collaboration} {et~al.}(2016{\natexlab{b}}){Gaia
  Collaboration}, {Prusti}, {de Bruijne}, {Brown}, {Vallenari}, {Babusiaux},
  {Bailer-Jones}, {Bastian}, {Biermann}, {Evans}, \&
  et~al.}]{GaiaCollaboration2016a}
{Gaia Collaboration}, {Prusti}, T., {de Bruijne}, J.~H.~J., {et~al.}
  2016{\natexlab{b}}, \aap, 595, A1

\bibitem[{{Gonz{\'a}lez Hern{\'a}ndez} {et~al.}(2008){Gonz{\'a}lez
  Hern{\'a}ndez}, {Bonifacio}, {Ludwig}, {Caffau}, {Spite}, {Spite}, {Cayrel},
  {Molaro}, {Hill}, {Fran{\c c}ois}, {Plez}, {Beers}, {Sivarani}, {Andersen},
  {Barbuy}, {Depagne}, {Nordstr{\"o}m}, \& {Primas}}]{GonzalezHernandez2008}
{Gonz{\'a}lez Hern{\'a}ndez}, J.~I., {Bonifacio}, P., {Ludwig}, H.-G., {et~al.}
  2008, \aap, 480, 233

\bibitem[{{Grevesse} {et~al.}(1981){Grevesse}, {Biemont}, {Lowe}, \&
  {Hannaford}}]{Grevesse1981}
{Grevesse}, N., {Biemont}, E., {Lowe}, R.~M., \& {Hannaford}, P. 1981, in Liege
  International Astrophysical Colloquia, Vol.~23, Liege International
  Astrophysical Colloquia, 211--222

\bibitem[{{Grevesse} {et~al.}(1989){Grevesse}, {Blackwell}, \&
  {Petford}}]{Grevesse1989}
{Grevesse}, N., {Blackwell}, D.~E., \& {Petford}, A.~D. 1989, \aap, 208, 157

\bibitem[{{Grevesse} {et~al.}(2015){Grevesse}, {Scott}, {Asplund}, \&
  {Sauval}}]{Grevesse2015}
{Grevesse}, N., {Scott}, P., {Asplund}, M., \& {Sauval}, A.~J. 2015, \aap, 573,
  A27

\bibitem[{{Hampel} {et~al.}(2016){Hampel}, {Stancliffe}, {Lugaro}, \&
  {Meyer}}]{Hampel2016}
{Hampel}, M., {Stancliffe}, R.~J., {Lugaro}, M., \& {Meyer}, B.~S. 2016, \apj,
  831, 171

\bibitem[{{Hansen} {et~al.}(2014){Hansen}, {Hansen}, {Christlieb}, {Yong},
  {Bessell}, {Garc{\'{\i}}a P{\'e}rez}, {Beers}, {Placco}, {Frebel}, {Norris},
  \& {Asplund}}]{Hansen2014}
{Hansen}, T., {Hansen}, C.~J., {Christlieb}, N., {et~al.} 2014, \apj, 787, 162

\bibitem[{{Hansen} {et~al.}(2016){Hansen}, {Andersen}, {Nordstr{\"o}m},
  {Beers}, {Placco}, {Yoon}, \& {Buchhave}}]{Hansen2016}
{Hansen}, T.~T., {Andersen}, J., {Nordstr{\"o}m}, B., {et~al.} 2016, \aap, 588,
  A3

\bibitem[{{Henden} {et~al.}(2016){Henden}, {Templeton}, {Terrell}, {Smith},
  {Levine}, \& {Welch}}]{Henden2016}
{Henden}, A.~A., {Templeton}, M., {Terrell}, D., {et~al.} 2016, VizieR Online
  Data Catalog, 2336

\bibitem[{{Ivans} {et~al.}(2006){Ivans}, {Simmerer}, {Sneden}, {Lawler},
  {Cowan}, {Gallino}, \& {Bisterzo}}]{Ivans2006}
{Ivans}, I.~I., {Simmerer}, J., {Sneden}, C., {et~al.} 2006, \apj, 645, 613

\bibitem[{{Jacobson} {et~al.}(2015){Jacobson}, {Keller}, {Frebel}, {Casey},
  {Asplund}, {Bessell}, {Da Costa}, {Lind}, {Marino}, {Norris}, {Pe{\~n}a},
  {Schmidt}, {Tisserand}, {Walsh}, {Yong}, \& {Yu}}]{Jacobson2015}
{Jacobson}, H.~R., {Keller}, S., {Frebel}, A., {et~al.} 2015, \apj, 807, 171

\bibitem[{{Jeon} {et~al.}(2017){Jeon}, {Besla}, \& {Bromm}}]{Jeon2017}
{Jeon}, M., {Besla}, G., \& {Bromm}, V. 2017, ArXiv e-prints, arXiv:1702.07355

\bibitem[{{Jordi} {et~al.}(2006){Jordi}, {Grebel}, \& {Ammon}}]{Jordi2006}
{Jordi}, K., {Grebel}, E.~K., \& {Ammon}, K. 2006, \aap, 460, 339

\bibitem[{{Kim} {et~al.}(2002){Kim}, {Demarque}, {Yi}, \&
  {Alexander}}]{Kim2002}
{Kim}, Y.-C., {Demarque}, P., {Yi}, S.~K., \& {Alexander}, D.~R. 2002, \apjs,
  143, 499

\bibitem[{{Kobayashi} {et~al.}(2006){Kobayashi}, {Umeda}, {Nomoto}, {Tominaga},
  \& {Ohkubo}}]{Kobayashi2006}
{Kobayashi}, C., {Umeda}, H., {Nomoto}, K., {Tominaga}, N., \& {Ohkubo}, T.
  2006, \apj, 653, 1145

\bibitem[{{Lawler} \& {Dakin}(1989)}]{Lawler1989}
{Lawler}, J.~E., \& {Dakin}, J.~T. 1989, Journal of the Optical Society of
  America B Optical Physics, 6, 1457

\bibitem[{{Lawler} {et~al.}(2013){Lawler}, {Guzman}, {Wood}, {Sneden}, \&
  {Cowan}}]{Lawler2013}
{Lawler}, J.~E., {Guzman}, A., {Wood}, M.~P., {Sneden}, C., \& {Cowan}, J.~J.
  2013, \apjs, 205, 11

\bibitem[{{Lee} {et~al.}(2008{\natexlab{a}}){Lee}, {Beers}, {Sivarani},
  {Allende Prieto}, {Koesterke}, {Wilhelm}, {Re Fiorentin}, {Bailer-Jones},
  {Norris}, {Rockosi}, {Yanny}, {Newberg}, {Covey}, {Zhang}, \&
  {Luo}}]{Lee2008}
{Lee}, Y.~S., {Beers}, T.~C., {Sivarani}, T., {et~al.} 2008{\natexlab{a}}, \aj,
  136, 2022

\bibitem[{{Lee} {et~al.}(2008{\natexlab{b}}){Lee}, {Beers}, {Sivarani},
  {Johnson}, {An}, {Wilhelm}, {Allende Prieto}, {Koesterke}, {Re Fiorentin},
  {Bailer-Jones}, {Norris}, {Yanny}, {Rockosi}, {Newberg}, {Cudworth}, \&
  {Pan}}]{Lee2008a}
---. 2008{\natexlab{b}}, \aj, 136, 2050

\bibitem[{{Li} {et~al.}(2015){Li}, {Aoki}, {Zhao}, {Honda}, {Christlieb}, \&
  {Suda}}]{Li2015a}
{Li}, H., {Aoki}, W., {Zhao}, G., {et~al.} 2015, \pasj, 67, 84

\bibitem[{{Li} {et~al.}(2010){Li}, {Christlieb}, {Sch{\"o}rck}, {Norris},
  {Bessell}, {Yong}, {Beers}, {Lee}, {Frebel}, \& {Zhao}}]{Li2010}
{Li}, H.~N., {Christlieb}, N., {Sch{\"o}rck}, T., {et~al.} 2010, \aap, 521, A10

\bibitem[{{Lind} {et~al.}(2009){Lind}, {Asplund}, \& {Barklem}}]{Lind2009}
{Lind}, K., {Asplund}, M., \& {Barklem}, P.~S. 2009, \aap, 503, 541

\bibitem[{{Lind} {et~al.}(2012){Lind}, {Bergemann}, \& {Asplund}}]{Lind2012}
{Lind}, K., {Bergemann}, M., \& {Asplund}, M. 2012, \mnras, 427, 50

\bibitem[{{Madau} \& {Dickinson}(2014)}]{Madau2014}
{Madau}, P., \& {Dickinson}, M. 2014, \araa, 52, 415

\bibitem[{{Martin} {et~al.}(1988){Martin}, {Fuhr}, \& {Wiese}}]{Martin1988}
{Martin}, G.~A., {Fuhr}, J.~R., \& {Wiese}, W.~L. 1988, {Atomic transition
  probabilities. Scandium through Manganese}

\bibitem[{{Masseron} {et~al.}(2012){Masseron}, {Johnson}, {Lucatello},
  {Karakas}, {Plez}, {Beers}, \& {Christlieb}}]{Masseron2012}
{Masseron}, T., {Johnson}, J.~A., {Lucatello}, S., {et~al.} 2012, \apj, 751, 14

\bibitem[{{Masseron} {et~al.}(2014){Masseron}, {Plez}, {Van Eck}, {Colin},
  {Daoutidis}, {Godefroid}, {Coheur}, {Bernath}, {Jorissen}, \&
  {Christlieb}}]{Masseron2014}
{Masseron}, T., {Plez}, B., {Van Eck}, S., {et~al.} 2014, \aap, 571, A47

\bibitem[{Matsuno {et~al.}(2017)Matsuno, Aoki, Suda, \& Li}]{Matsuno2017}
Matsuno, T., Aoki, W., Suda, T., \& Li, H. 2017, \pasj, 69, 24

\bibitem[{{Mel{\'e}ndez} \& {Barbuy}(2009)}]{Melendez2009}
{Mel{\'e}ndez}, J., \& {Barbuy}, B. 2009, \aap, 497, 611

\bibitem[{{Mel{\'e}ndez} {et~al.}(2010){Mel{\'e}ndez}, {Casagrande},
  {Ram{\'{\i}}rez}, {Asplund}, \& {Schuster}}]{Melendez2010}
{Mel{\'e}ndez}, J., {Casagrande}, L., {Ram{\'{\i}}rez}, I., {Asplund}, M., \&
  {Schuster}, W.~J. 2010, \aap, 515, L3

\bibitem[{{Michaud} {et~al.}(2015){Michaud}, {Alecian}, \&
  {Richer}}]{Michaud2015}
{Michaud}, G., {Alecian}, G., \& {Richer}, J. 2015, {Atomic Diffusion in Stars}
  (Springer International Publishing), doi:10.1007/978-3-319-19854-5

\bibitem[{{Moity}(1983)}]{Moity1983}
{Moity}, J. 1983, \aaps, 52, 37

\bibitem[{{Morton}(1991)}]{Morton1991}
{Morton}, D.~C. 1991, \apjs, 77, 119

\bibitem[{{Nissen} {et~al.}(2007){Nissen}, {Akerman}, {Asplund}, {Fabbian},
  {Kerber}, {Kaufl}, \& {Pettini}}]{Nissen2007}
{Nissen}, P.~E., {Akerman}, C., {Asplund}, M., {et~al.} 2007, \aap, 469, 319

\bibitem[{{Noguchi} {et~al.}(2002){Noguchi}, {Aoki}, {Kawanomoto}, {Ando},
  {Honda}, {Izumiura}, {Kambe}, {Okita}, {Sadakane}, {Sato}, {Tajitsu},
  {Takada-Hidai}, {Tanaka}, {Watanabe}, \& {Yoshida}}]{Noguchi2002}
{Noguchi}, K., {Aoki}, W., {Kawanomoto}, S., {et~al.} 2002, \pasj, 54, 855

\bibitem[{{Nomoto} {et~al.}(2013){Nomoto}, {Kobayashi}, \&
  {Tominaga}}]{Nomoto2013}
{Nomoto}, K., {Kobayashi}, C., \& {Tominaga}, N. 2013, \araa, 51, 457

\bibitem[{{Norris} {et~al.}(2000){Norris}, {Beers}, \& {Ryan}}]{Norris2000}
{Norris}, J.~E., {Beers}, T.~C., \& {Ryan}, S.~G. 2000, \apj, 540, 456

\bibitem[{{Norris} {et~al.}(2013{\natexlab{a}}){Norris}, {Bessell}, {Yong},
  {Christlieb}, {Barklem}, {Asplund}, {Murphy}, {Beers}, {Frebel}, \&
  {Ryan}}]{Norris2013}
{Norris}, J.~E., {Bessell}, M.~S., {Yong}, D., {et~al.} 2013{\natexlab{a}},
  \apj, 762, 25

\bibitem[{{Norris} {et~al.}(2013{\natexlab{b}}){Norris}, {Yong}, {Bessell},
  {Christlieb}, {Asplund}, {Gilmore}, {Wyse}, {Beers}, {Barklem}, {Frebel}, \&
  {Ryan}}]{Norris2013a}
{Norris}, J.~E., {Yong}, D., {Bessell}, M.~S., {et~al.} 2013{\natexlab{b}},
  \apj, 762, 28

\bibitem[{{O'Brian} {et~al.}(1991){O'Brian}, {Wickliffe}, {Lawler}, {Whaling},
  \& {Brault}}]{OBrian1991}
{O'Brian}, T.~R., {Wickliffe}, M.~E., {Lawler}, J.~E., {Whaling}, W., \&
  {Brault}, J.~W. 1991, Journal of the Optical Society of America B Optical
  Physics, 8, 1185

\bibitem[{{Piau} {et~al.}(2006){Piau}, {Beers}, {Balsara}, {Sivarani},
  {Truran}, \& {Ferguson}}]{Piau2006}
{Piau}, L., {Beers}, T.~C., {Balsara}, D.~S., {et~al.} 2006, \apj, 653, 300

\bibitem[{{Pickering} {et~al.}(2001){Pickering}, {Thorne}, \&
  {Perez}}]{Pickering2001}
{Pickering}, J.~C., {Thorne}, A.~P., \& {Perez}, R. 2001, \apjs, 132, 403

\bibitem[{{Pinnington} {et~al.}(1995){Pinnington}, {Berends}, \&
  {Lumsden}}]{Pinnington1995}
{Pinnington}, E.~H., {Berends}, R.~W., \& {Lumsden}, M. 1995, Journal of
  Physics B Atomic Molecular Physics, 28, 2095

\bibitem[{{Piskunov} {et~al.}(1995){Piskunov}, {Kupka}, {Ryabchikova}, {Weiss},
  \& {Jeffery}}]{Piskunov1995}
{Piskunov}, N.~E., {Kupka}, F., {Ryabchikova}, T.~A., {Weiss}, W.~W., \&
  {Jeffery}, C.~S. 1995, \aaps, 112, 525

\bibitem[{{Placco} {et~al.}(2016){Placco}, {Beers}, {Reggiani}, \&
  {Mel{\'e}ndez}}]{Placco2016}
{Placco}, V.~M., {Beers}, T.~C., {Reggiani}, H., \& {Mel{\'e}ndez}, J. 2016,
  \apjl, 829, L24

\bibitem[{{Planck Collaboration} {et~al.}(2016){Planck Collaboration}, {Adam},
  {Ade}, {Aghanim}, {Akrami}, {Alves}, {Arg{\"u}eso}, {Arnaud}, {Arroja},
  {Ashdown}, \& et~al.}]{PlanckCollaboration2016}
{Planck Collaboration}, {Adam}, R., {Ade}, P.~A.~R., {et~al.} 2016, \aap, 594,
  A1

\bibitem[{{Reggiani} {et~al.}(2016){Reggiani}, {Mel{\'e}ndez}, {Yong},
  {Ram{\'{\i}}rez}, \& {Asplund}}]{Reggiani2016}
{Reggiani}, H., {Mel{\'e}ndez}, J., {Yong}, D., {Ram{\'{\i}}rez}, I., \&
  {Asplund}, M. 2016, \aap, 586, A67

\bibitem[{{Richard} {et~al.}(2005){Richard}, {Michaud}, \&
  {Richer}}]{Richard2005}
{Richard}, O., {Michaud}, G., \& {Richer}, J. 2005, \apj, 619, 538

\bibitem[{{Ryabchikova} {et~al.}(1994){Ryabchikova}, {Hill}, {Landstreet},
  {Piskunov}, \& {Sigut}}]{Ryabchikova1994}
{Ryabchikova}, T.~A., {Hill}, G.~M., {Landstreet}, J.~D., {Piskunov}, N., \&
  {Sigut}, T.~A.~A. 1994, \mnras, 267, 697

\bibitem[{{Ryan} {et~al.}(1996){Ryan}, {Beers}, {Deliyannis}, \&
  {Thorburn}}]{Ryan1996}
{Ryan}, S.~G., {Beers}, T.~C., {Deliyannis}, C.~P., \& {Thorburn}, J.~A. 1996,
  \apj, 458, 543

\bibitem[{{Ryan} {et~al.}(1999){Ryan}, {Norris}, \& {Beers}}]{Ryan1999}
{Ryan}, S.~G., {Norris}, J.~E., \& {Beers}, T.~C. 1999, \apj, 523, 654

\bibitem[{{Sbordone} {et~al.}(2010){Sbordone}, {Bonifacio}, {Caffau}, {Ludwig},
  {Behara}, {Gonz{\'a}lez Hern{\'a}ndez}, {Steffen}, {Cayrel}, {Freytag},
  {van't Veer}, {Molaro}, {Plez}, {Sivarani}, {Spite}, {Spite}, {Beers},
  {Christlieb}, {Fran{\c c}ois}, \& {Hill}}]{Sbordone2010}
{Sbordone}, L., {Bonifacio}, P., {Caffau}, E., {et~al.} 2010, \aap, 522, A26

\bibitem[{{Schlafly} \& {Finkbeiner}(2011)}]{Schlafly2011}
{Schlafly}, E.~F., \& {Finkbeiner}, D.~P. 2011, \apj, 737, 103

\bibitem[{{Sch{\"o}rck} {et~al.}(2009){Sch{\"o}rck}, {Christlieb}, {Cohen},
  {Beers}, {Shectman}, {Thompson}, {McWilliam}, {Bessell}, {Norris},
  {Mel{\'e}ndez}, {Ram{\'{\i}}rez}, {Haynes}, {Cass}, {Hartley}, {Russell},
  {Watson}, {Zickgraf}, {Behnke}, {Fechner}, {Fuhrmeister}, {Barklem},
  {Edvardsson}, {Frebel}, {Wisotzki}, \& {Reimers}}]{Schoerck2009}
{Sch{\"o}rck}, T., {Christlieb}, N., {Cohen}, J.~G., {et~al.} 2009, \aap, 507,
  817

\bibitem[{{Shimazaki} \& {Shinomoto}(2007)}]{Shimazaki2007}
{Shimazaki}, H., \& {Shinomoto}, S. 2007, Neural Computation, 19, 1503

\bibitem[{{Smith} \& {Raggett}(1981)}]{Smith1981}
{Smith}, G., \& {Raggett}, D.~S.~J. 1981, Journal of Physics B Atomic Molecular
  Physics, 14, 4015

\bibitem[{{Smith} {et~al.}(1998){Smith}, {Lambert}, \& {Nissen}}]{Smith1998}
{Smith}, V.~V., {Lambert}, D.~L., \& {Nissen}, P.~E. 1998, \apj, 506, 405

\bibitem[{{Sobeck} {et~al.}(2007){Sobeck}, {Lawler}, \& {Sneden}}]{Sobeck2007}
{Sobeck}, J.~S., {Lawler}, J.~E., \& {Sneden}, C. 2007, \apj, 667, 1267

\bibitem[{{Spite} \& {Spite}(1982{\natexlab{a}})}]{Spite1982}
{Spite}, F., \& {Spite}, M. 1982{\natexlab{a}}, \aap, 115, 357

\bibitem[{{Spite} \& {Spite}(1982{\natexlab{b}})}]{Spite1982a}
{Spite}, M., \& {Spite}, F. 1982{\natexlab{b}}, \nat, 297, 483

\bibitem[{{Spite} {et~al.}(2006){Spite}, {Cayrel}, {Hill}, {Spite}, {Fran{\c
  c}ois}, {Plez}, {Bonifacio}, {Molaro}, {Depagne}, {Andersen}, {Barbuy},
  {Beers}, {Nordstr{\"o}m}, \& {Primas}}]{Spite2006}
{Spite}, M., {Cayrel}, R., {Hill}, V., {et~al.} 2006, \aap, 455, 291

\bibitem[{{Stancliffe} {et~al.}(2007){Stancliffe}, {Glebbeek}, {Izzard}, \&
  {Pols}}]{Stancliffe2007}
{Stancliffe}, R.~J., {Glebbeek}, E., {Izzard}, R.~G., \& {Pols}, O.~R. 2007,
  \aap, 464, L57

\bibitem[{{Starkenburg} {et~al.}(2014){Starkenburg}, {Shetrone}, {McConnachie},
  \& {Venn}}]{Starkenburg2014}
{Starkenburg}, E., {Shetrone}, M.~D., {McConnachie}, A.~W., \& {Venn}, K.~A.
  2014, \mnras, 441, 1217

\bibitem[{{Suda} {et~al.}(2011){Suda}, {Yamada}, {Katsuta}, {Komiya},
  {Ishizuka}, {Aoki}, \& {Fujimoto}}]{Suda2011}
{Suda}, T., {Yamada}, S., {Katsuta}, Y., {et~al.} 2011, \mnras, 412, 843

\bibitem[{{Suda} {et~al.}(2008){Suda}, {Katsuta}, {Yamada}, {Suwa}, {Ishizuka},
  {Komiya}, {Sorai}, {Aikawa}, \& {Fujimoto}}]{Suda2008}
{Suda}, T., {Katsuta}, Y., {Yamada}, S., {et~al.} 2008, \pasj, 60, 1159

\bibitem[{{Tominaga} {et~al.}(2014){Tominaga}, {Iwamoto}, \&
  {Nomoto}}]{Tominaga2014}
{Tominaga}, N., {Iwamoto}, N., \& {Nomoto}, K. 2014, \apj, 785, 98

\bibitem[{{Wiese} \& {Martin}(1980)}]{Wiese1980}
{Wiese}, W.~L., \& {Martin}, G.~A. 1980, {Wavelengths and transition
  probabilities for atoms and atomic ions: Part 2. Transition probabilities}

\bibitem[{{Wood} {et~al.}(2013){Wood}, {Lawler}, {Sneden}, \&
  {Cowan}}]{Wood2013}
{Wood}, M.~P., {Lawler}, J.~E., {Sneden}, C., \& {Cowan}, J.~J. 2013, \apjs,
  208, 27

\bibitem[{{Yamada} {et~al.}(2013){Yamada}, {Suda}, {Komiya}, {Aoki}, \&
  {Fujimoto}}]{Yamada2013}
{Yamada}, S., {Suda}, T., {Komiya}, Y., {Aoki}, W., \& {Fujimoto}, M.~Y. 2013,
  \mnras, 436, 1362

\bibitem[{{Yanny} {et~al.}(2009){Yanny}, {Rockosi}, {Newberg}, {Knapp},
  {Adelman-McCarthy}, {Alcorn}, {Allam}, {Allende Prieto}, {An}, {Anderson},
  {Anderson}, {Bailer-Jones}, {Bastian}, {Beers}, {Bell}, {Belokurov},
  {Bizyaev}, {Blythe}, {Bochanski}, {Boroski}, {Brinchmann}, {Brinkmann},
  {Brewington}, {Carey}, {Cudworth}, {Evans}, {Evans}, {Gates}, {G{\"a}nsicke},
  {Gillespie}, {Gilmore}, {Nebot Gomez-Moran}, {Grebel}, {Greenwell}, {Gunn},
  {Jordan}, {Jordan}, {Harding}, {Harris}, {Hendry}, {Holder}, {Ivans},
  {Ivezi{\v c}}, {Jester}, {Johnson}, {Kent}, {Kleinman}, {Kniazev},
  {Krzesinski}, {Kron}, {Kuropatkin}, {Lebedeva}, {Lee}, {French Leger},
  {L{\'e}pine}, {Levine}, {Lin}, {Long}, {Loomis}, {Lupton}, {Malanushenko},
  {Malanushenko}, {Margon}, {Martinez-Delgado}, {McGehee}, {Monet}, {Morrison},
  {Munn}, {Neilsen}, {Nitta}, {Norris}, {Oravetz}, {Owen}, {Padmanabhan},
  {Pan}, {Peterson}, {Pier}, {Platson}, {Re Fiorentin}, {Richards}, {Rix},
  {Schlegel}, {Schneider}, {Schreiber}, {Schwope}, {Sibley}, {Simmons},
  {Snedden}, {Allyn Smith}, {Stark}, {Stauffer}, {Steinmetz}, {Stoughton},
  {SubbaRao}, {Szalay}, {Szkody}, {Thakar}, {Sivarani}, {Tucker}, {Uomoto},
  {Vanden Berk}, {Vidrih}, {Wadadekar}, {Watters}, {Wilhelm}, {Wyse}, {Yarger},
  \& {Zucker}}]{Yanny2009}
{Yanny}, B., {Rockosi}, C., {Newberg}, H.~J., {et~al.} 2009, \aj, 137, 4377

\bibitem[{{Yong} {et~al.}(2013){Yong}, {Norris}, {Bessell}, {Christlieb},
  {Asplund}, {Beers}, {Barklem}, {Frebel}, \& {Ryan}}]{Yong2013a}
{Yong}, D., {Norris}, J.~E., {Bessell}, M.~S., {et~al.} 2013, \apj, 762, 27

\bibitem[{{York} {et~al.}(2000){York}, {Adelman}, {Anderson}, {Anderson},
  {Annis}, {Bahcall}, {Bakken}, {Barkhouser}, {Bastian}, {Berman}, {Boroski},
  {Bracker}, {Briegel}, {Briggs}, {Brinkmann}, {Brunner}, {Burles}, {Carey},
  {Carr}, {Castander}, {Chen}, {Colestock}, {Connolly}, {Crocker}, {Csabai},
  {Czarapata}, {Davis}, {Doi}, {Dombeck}, {Eisenstein}, {Ellman}, {Elms},
  {Evans}, {Fan}, {Federwitz}, {Fiscelli}, {Friedman}, {Frieman}, {Fukugita},
  {Gillespie}, {Gunn}, {Gurbani}, {de Haas}, {Haldeman}, {Harris}, {Hayes},
  {Heckman}, {Hennessy}, {Hindsley}, {Holm}, {Holmgren}, {Huang}, {Hull},
  {Husby}, {Ichikawa}, {Ichikawa}, {Ivezi{\'c}}, {Kent}, {Kim}, {Kinney},
  {Klaene}, {Kleinman}, {Kleinman}, {Knapp}, {Korienek}, {Kron}, {Kunszt},
  {Lamb}, {Lee}, {Leger}, {Limmongkol}, {Lindenmeyer}, {Long}, {Loomis},
  {Loveday}, {Lucinio}, {Lupton}, {MacKinnon}, {Mannery}, {Mantsch}, {Margon},
  {McGehee}, {McKay}, {Meiksin}, {Merelli}, {Monet}, {Munn}, {Narayanan},
  {Nash}, {Neilsen}, {Neswold}, {Newberg}, {Nichol}, {Nicinski}, {Nonino},
  {Okada}, {Okamura}, {Ostriker}, {Owen}, {Pauls}, {Peoples}, {Peterson},
  {Petravick}, {Pier}, {Pope}, {Pordes}, {Prosapio}, {Rechenmacher}, {Quinn},
  {Richards}, {Richmond}, {Rivetta}, {Rockosi}, {Ruthmansdorfer}, {Sandford},
  {Schlegel}, {Schneider}, {Sekiguchi}, {Sergey}, {Shimasaku}, {Siegmund},
  {Smee}, {Smith}, {Snedden}, {Stone}, {Stoughton}, {Strauss}, {Stubbs},
  {SubbaRao}, {Szalay}, {Szapudi}, {Szokoly}, {Thakar}, {Tremonti}, {Tucker},
  {Uomoto}, {Vanden Berk}, {Vogeley}, {Waddell}, {Wang}, {Watanabe},
  {Weinberg}, {Yanny}, {Yasuda}, \& {SDSS Collaboration}}]{York2000}
{York}, D.~G., {Adelman}, J., {Anderson}, Jr., J.~E., {et~al.} 2000, \aj, 120,
  1579

\end{thebibliography}
\end{document}